\documentclass[paper]{geophysics}
\usepackage{amsmath}

\begin{document}

\title{Multi-Objective Bayesian Optimisation \\and Joint Inversion for Active Sensor Fusion}



\author{Sebastian Haan\footnotemark[1], Fabio Ramos \footnotemark[2] and  R. Dietmar M\"uller \footnotemark[3]}

\lefthead{Haan, Ramos, M\"uller}
\righthead{Multi-Objective Bayesian Optimisation}

\maketitle

\footnote{The University of Sydney, Sydney Informatics Hub, Sydney, NSW, 2008, Australia. \\ E-mail:  sebastian.haan@sydney.edu.au (corresponding author)\\
\footnotemark[2]The University of Sydney,  School of Computer Science, Sydney, NSW, 2006, Australia.\\  Email: fabio.ramos@sydney.edu.au \\
\footnotemark[3]The University of Sydney,  School of Geosciences, Sydney, NSW, 2006, Australia. \\ Email: dietmar.muller@sydney.edu.au}


\begin{abstract}
A critical decision process in data acquisition for mineral and energy resource exploration is how to efficiently combine a variety of sensor types and to minimize total cost. We propose a probabilistic framework for multi-objective optimisation and inverse problems given an expensive cost function for allocating new measurements. This new method is devised to jointly solve multi-linear forward models of 2D-sensor data and 3D-geophysical properties using sparse Gaussian Process kernels while taking into account the cross-variances of different parameters. Multiple optimisation strategies are tested and evaluated on a set of synthetic and real geophysical data. We demonstrate the advantages on a specific example of a joint inverse problem, recommending where to place new drill-core measurements given 2D gravity and magnetic sensor data, the same approach can be applied to a variety of remote sensing problems with linear forward models - ranging from constraints limiting surface access for data acquisition to adaptive multi-sensor positioning.
\end{abstract}

\section{Introduction}

One of the most important aspects when dealing with incomplete information is where to collect new observations. This may include drill-site placement or the layout of a particular geophysical survey, in particular if measurements are very costly or resources are limited. Bayesian Optimisation (BO) solves this decision-making problem by finding global optimal solutions using a probabilistic framework given multiple measurements, prior knowledge, and model uncertainties. This offers solutions for wide range
of sensor fusion problems and allocation problems such as:
What is the optimal mixture and placement of different sensors?
Where to sample if the cost function is incomplete or if
there are large uncertainties in future total budget allocations?
How to position sensor grids over time if the model state is
dynamic or has a moving target?

\par

Typically the function(s) involved in BO are unknown or very expensive to evaluate, and are therefore surrogated by prior models, such as Gaussian process (GP) models for tractability \citep[see][ for an introduction to GPs]{rasmussen2006gaussian}. Once new data are available to evaluate the objective function, the prior is updated to form the posterior distribution over the function space. \par

While BO research began by focusing on single-objective bound-constrained optimisation \citep[see for earliest work][]{kushner1964, mockus1975}, e.g., using the Expected Improvement (EI) criterion \citep{mockus1975bayesian,jones1998}, it has rapidly developed to address multi-objective problems \citep[see, e.g.,][]{emmerich2005,jeong2005}. Many criteria have been proposed for finding the optimal placements of sensors, where the goal is to maximize the cumulative reward for placing sensors by optimally balancing exploration and exploitation with as few evaluations: from the multi-armed bandit paradigm \citep{robbins1952, srinivas2009} to maximizing the information gain and the mutual information between the chosen and unselected locations \citep{caselton1984, krause2008}, and other value of information criteria \citep[see, e.g.][]{eidsvik2015, mukerji2001, osborne2008}. Today multi-Objective BO is recognised as a robust statistical method in handling multiple constraints and to find the optimal solution for competing objectives and multi-task problems.\par

 The set of objectives is defined in an acquisition function. While there exist many approaches in the literature to define and optimize acquisition functions \citep[for an overview see, e.g.,][]{brochu2010tutorial}, the main challenge still lies in constructing a prior model that can be computationally efficiently evaluated and that transforms the combined data and uncertainties from multiple sensor measurements coherently into a posterior function approximating the true model, such as a 3D model of a portion of the continental crust; this is also known as an inverse problem.\par

In geology and geophysics, inverse problems occur whenever the goal is to reconstruct the geological conditions, i.e. the distribution of physical rock properties, that give rise to a set of geophysical observations. Since the number of possible geological configurations is typically greater than the number of observational constraints, the problem is nearly always under-determined. A typical example is gravity anomalies remotely measured at the Earth's surface, which reflect the 3D density distribution within the Earth. A given gravity anomaly at the surface may be caused by a given density anomaly near the surface or by a much larger density anomaly more deeply buried. Surface gravity anomaly data cannot distinguish between such alternative scenarios, but may provide constraints on geological conditions if certain assumptions are provided, such as a maximal and minimal density variation from the mean density. One important tool to  solve the inverse problem are Monte Carlo (MC) methods \citep[see for a review ][]{mosegaard2002}, which provide a probabilistic approach by obtaining a posterior distribution of possible solutions. MC methods are in particular useful when no analytical forward model is available or if linearization of the relation between data and model is not possible (e.g. for seismic wave models). However, MC methods are computational very expensive and typically require  prior defined knowledge about the underlying geological structure and fine-tuned parameter distribution. \par

Another promising way to infer the geophysical properties is by simultaneously interpreting multiple sensor measurements using a single model. The motivation behind this joint inversion is to provide a better constrained joint solution of multiple, and often distinct sensor types, rather than taking individual solutions that only satisfy their aspect of data on their own. In case these multiple sensors are sensitive to different aspects of the geology, e.g. density and magnetic susceptibility, we can take advantage of the statistical properties of a model that simultaneously need to satisfy two or more independent measurements. The resulting 'boost' in information is dependent on the strength of the correlation between the different physical properties that are measured, but if statistically properly taken  into account, this approach can offer a much greater value than the 'sum of their parts'. \par

Multiple methods have been developed to tackle the joint inverse problems: \cite{Bosch2006} have used Monte Carlo methods for joint gravity and magnetic 3D inversion. \cite{fedi1993} apply a 3D method that provides gravity and magnetic field information along the vertical direction for retrieving deep sources. More recently, 3D stochastic joint inversion of gravity and magnetic data has been developed by \cite{shamsipour2012}, which is based on cokriging and applied to estimate density and magnetic susceptibility distributions from field data. An alternative approach is to use a GP, which is a powerful non-parametric Bayesian method to solve joint inverse problems \citep{reid2013bayesian}. The main practical advantage of a probabilistic (Bayesian) approach is that predictions are described by posterior distributions for each location, which quantify the uncertainty in the predictions and their credible intervals. Most other work in the field of joint inversion only report point estimates of the quantities of interest \citep[see, e.g.,][]{zeyen1993, gallardo2003, moorkamp2011}. 

In this paper we propose a probabilistic framework that combines coherent joint inverse problems and multi-objective optimisation given an expensive cost function. This technique is applied to jointly solve multi-linear forward models of distinct sensor data to 3D-geophysical properties using sparse Gaussian Process kernels while taking into account the correlations between distinct geological aspects. In general, the method provides a very flexible approach for a large range of sensor fusion problems and is computationally efficient enough to rapidly update a model with new data while solving multiple measurement decision problems simultaneously. Our data driven approach does not attempt to include prior knowledge of any geological structure and returns the probability distribution of 3D reconstructed geophysical properties based on sensor data input only. The resulting 3D cubes and visualizations provide a suitable basis to guide further exploration or to generate subsequent models using additional geological constrains. \par

The outline of this paper is as follows. Section 2 describes the underlying probabilistic framework and computational methods for geological inversion and strategies for Bayesian optimisation. The method is then evaluated on a set of synthetic data as described in Section 3 and tested on real case scenario in Section 4. The results, limitations, and further applications are discussed in Section 5.


\section{Methodology} \label{sec_method}
\begin{figure}[!t]
\centering
\includegraphics[width=4.5in]{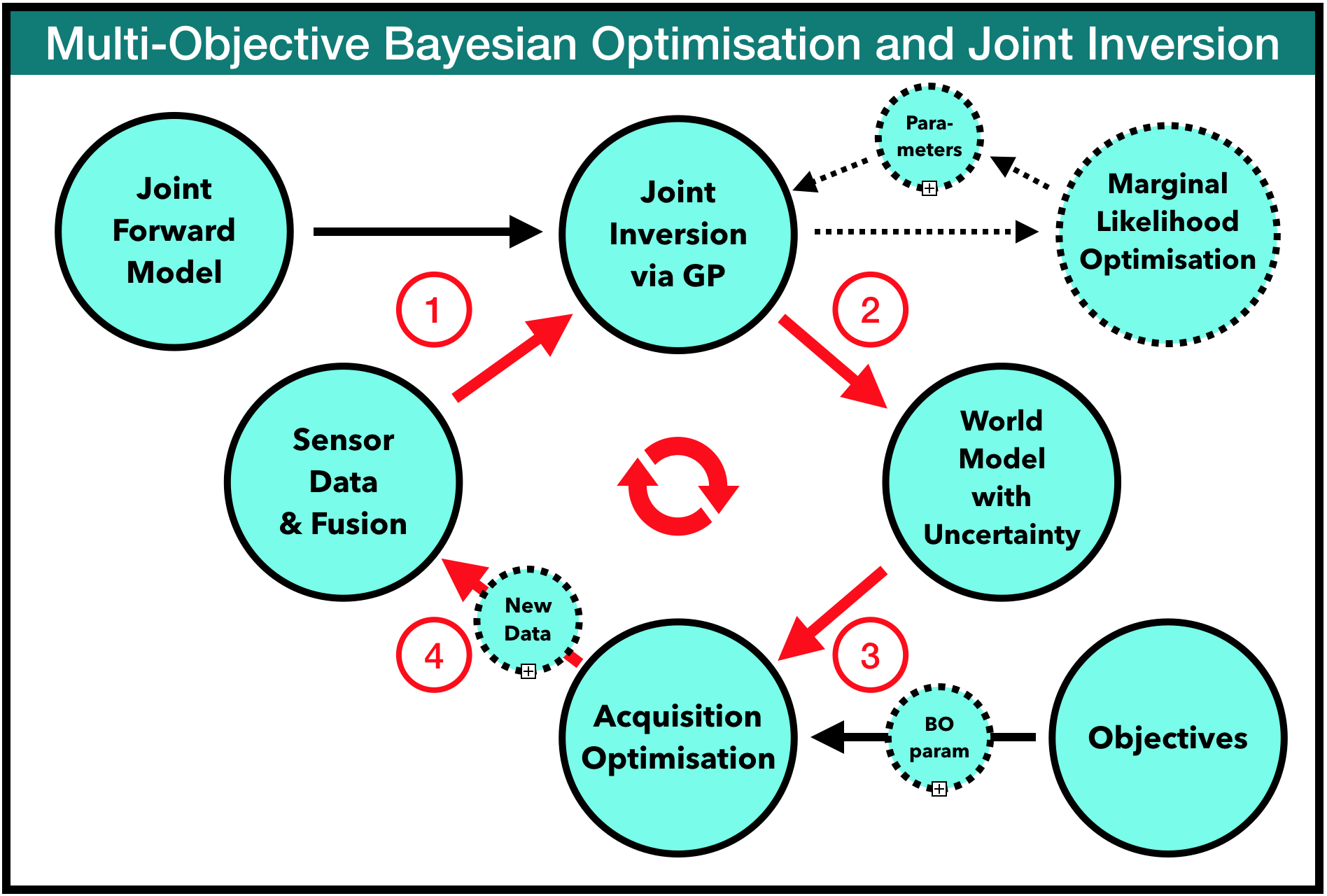}
\caption{Graphical model of the probabilistic framework for
multi-objective optimisation and joint inversion; the process
stages are: (1) Joint inversion via Gaussian Processes based
on multiple sensor data fusion and forward model, (2) Posterior
model generation of 3D multi-geophysical properties. (3)
maximization of acquisition function (objectives) to allocate optimal new
sensor location and type. (4) New acquired data are combined
with existing data; process repeats until maximum number of
iterations is achieved.}

\label{fig_graphmodel}
\end{figure}

Our approach for joint inversion and optimisation builds upon a fully probabilistic model, including likelihood objectives for model selection and complete uncertainty propagation through all stages of data processing: from prior selection and input, forward model inversion, to the final sensor optimisation and real world model output. Figure~\ref{fig_graphmodel} shows a graphical model of the developed probabilistic framework.

In general the relationship between a physical system (or  its model parameters) $\Phi$ and the observed sensor data $y$ is described by a forward model
\begin{equation}
y = G(\Phi)    
\end{equation}
given the transformation function $G$. For a linear system such as, e.g., the gravitational or magnetic field, the problem reduces to 
\begin{equation}
y = G \Phi  
\end{equation}
where $G$ is the transformation operator (often described as matrix).
To infer the underlying parameters of the physical system, the inverse operator, $G^{-1}$, is required. However, direct matrix inversion is only possible for a limited range of cases ($G$ must be a square matrix and non-singular, i.e $\det(|G|)\neq0$), and in most applications $G$ is not even a square matrix. To solve the inverse problem for the more general case, we deploy a Bayesian framework with GP priors over the physical properties $\Phi$ of interest (e.g., rock properties). An important advantage of the Bayesian method is that it generates a predictive distribution with a mean $\mu(\Phi)$ and variance $\sigma^2(\Phi)$, which are prerequisites for Bayesian optimisation with respect to, e.g., information gain from a new measurement.

\subsection{Linear Forward Models}
In this work we apply linear forward models to infer geological rock properties (i.e., density and magnetic susceptibility) in 3D (Latitude, Longitude, Depth) from 2D (Latitude, Longitude) sensor-mesh data, which are typically placed on top of the earth surface or from aerial surveys. 
The sensors measure the variation of the gravitational and magnetic field: the gravitational field variations are driven by the density variations of the material, while magnetic field variations are described by the superposition of Earth's magnetic field and the induced secondary fields caused by magnetisation of materials, which depends upon their magnetic susceptibility and is linear as first-order approximation (neglecting residual remanent magnetisation).\par 

In practice, the magnetic field information is reduced to the residual between natural and induced field --- defined as anomalous magnetic field component. The gravitational and magnetic forward model, $T_{grav}$ and $T_{magn}$ can be determined analytically by using Li's tractable approximation \citep{li1996, li1998} for a 3D field of prisms of constant susceptibility and density, and we apply this prism shape model  to compute the corresponding sensor sensitivity for gravity and anomalous magnetic field related to each prism cell \citep[see, e.g.,][]{nagy2000gravitational, reid2013bayesian}.

\subsection{Gaussian Processes}
GPs are a flexible, probabilistic approach using kernel machines with non-parametric priors, and are successfully used in a large range of machine learning problems \citep[see][ for more details]{rasmussen2006gaussian}. An important advantage of using GPs for data processing and machine learning is their ability to propagate consistently uncertainties from input to output space under the Bayesian formalism, which is an essential requirement for Bayesian optimisation.  A GP $f (\mathbf{x }) \sim \mathcal{ G } \mathcal{ P } \left( \mu ( \mathbf{ x } ) , k \left( \mathbf { x } , \mathbf { x } ^ { \prime } \right) \right)$ is completely determined by its mean $\mu (\mathbf{ x })  = \mathrm{ E } [f (\mathbf{ x })]$ and covariance $ k \left( \mathbf { x } , \mathbf { x } ^ { \prime } \right)  = \mathrm { E } \left[ ( f ( \mathbf { x } ) - \mu ( \mathbf { x } ) ) \left( f \left( \mathbf { x } ^ { \prime } \right) - \mu \left( \mathbf { x } ^ { \prime } \right) \right) \right] $, essentially placing a multivariate Gaussian distribution over the space of functions that map the input to the output, or informally, to measure the similarity between points as function of, e.g., their distance (the kernel function) and to predict a Gaussian distribution over $f(x^*)$ (i.e., a mean value and variance) at any new sampling location (unseen point) $x^*$ from training data. \par

GPs are mathematically closely related to many well-known models, including spline models, Bayesian linear models, and, to some extent, support vector machines and deep neural networks (under certain conditions, i.e., a certain covariance kernel and infinite hidden layers). Another advantage of GPs is, as for other kernel methods, that their marginal likelihood function is well defined by the values of their hyper-parameters, and can thus be optimized. This reasoning about functions under uncertainty and their well-tuned interpolation character allows GPs to work extremely well for sparse data, but can come with an expensive computational cost function for large data ($N>$10000) as it scales with $\mathcal{ O } (N^3)$. However, recent advances in approximate (variational) inference as well as using sparse covariance functions make GPs also applicable for big data problems by reducing the complexity to $\mathcal { O } (N^2)$ and $\mathcal { O } ( M \log M )$, respectively, where $M$ is the number of inducing points with $M<<N$.  For comparison, the computational time to train deep neural networks scales typically with $\mathcal { O } ( CN )$, where $C$ is the number of layers and $N$ the average number of perceptrons per layer.\par

\subsubsection*{Choice of Covariance Functions}
The choice for an appropriate covariance function is important, as the GP's output directly depends on it. There are many stationary (invariant to translation in input space) and non-stationary covariance functions \citep[for a review see][]{rasmussen2006gaussian}; the former are typically defined by radial basis functions (RBF) such as the squared-exponential covariance function,
\begin{equation}
k  \left( x  , x { \prime }  \right) = \sigma _{0}^{2} \exp \left( - \frac{ 1 } { 2 l^ { 2 } } \left( x  - x^{ \prime }  \right) ^ { 2 } \right) 
\end{equation}
which depends on the radial distance between points $x$ and $x^{ \prime } $ and at least two parameters, one defining the length-scale $l$ (multiple dimensions can have multiple length-scales) and the other the signal variance $\sigma_{0}^2$.
These parameters of the covariance function are referred to as the hyperparameters $\Theta$ of the GP, which can be either given by a fixed covariance scale and noise, or learned from data by optimising the marginal likelihood. The latter allows us also to marginalise over a range of GP hyperparameters such as the length-scale parameters $l$. Computationally this can be done by defining a grid of hyperparameter values and then weight the posterior distribution with the corresponding marginalised likelihood values. In general, it is advisable to compute the joint inversion at least for a range of $l$ parameters and to compare the reconstructed models against each other.\par 

To handle the computational problem of inverting a large covariance matrix, ~\cite{melkumyan2009sparse} proposed an intrinsically sparse covariance function $K(r,l,\sigma_0)=$
\begin{equation}
  \begin{cases}
    K(r,l,\sigma_0)= \sigma_0 \Big[ \dfrac{1}{3}(2+\cos{(2\pi\dfrac{r}{l})})(1-\dfrac{r}{l}) + \dfrac{1}{2\pi}\sin{(2\pi\dfrac{r}{l})} \Big], & \text{if $r<l$.}\\
    0, & \text{if $r\geq l $.}
  \end{cases}
\end{equation}
where $\sigma_0 > 0$ is a constant coefficient, $l > 0$ is a given scale
and $r$ is the distance between the points: $r = \left| x - x ^ { \prime } \right|$.

\subsubsection*{Multi-Kernel Covariance Functions}
To take fully into account cross-covariances between multiple model parameters (e.g., rock density and magnetic susceptibility), we construct cross-covariance terms between all kernel pairs. One important requirement for constructing cross-covariance terms is that they must be defined to be both positive semi-definite and informative; for an overview how to construct such as matrix in detail see \cite{melkumyan2011multi}. Unless explicitly mentioned, all subsequent analysis is based on sparse covariances. Following \cite{melkumyan2011multi}, sparse-sparse cross-covariance terms are defined as 
\begin{equation}
\begin{split}
k_{S_1 \times S_2} (r; l_1, l_2) = \dfrac{2}{3\sqrt{l_1 l_2}}
\Bigg[ l_{min} + \dfrac{1}{\pi} \dfrac{l_{max}^3}{l_{max}^2 - l_{min}^2} \\
\times \sin{\Big(\pi \dfrac{l_{min}}{l_{max}}\Big)} \cos{\Big(\dfrac{2\pi r}{l_{max}}\Big)}\Bigg] \text { if } r\leq \dfrac{|l_2 - l_1|}{2} \\
k_{S_1 \times S_2} (r; l_1, l_2) = \dfrac{2}{3\sqrt{l_1 l_2}}
\Bigg[ l_{mean} - r + \dfrac{l_1^3 \sin{\Big(\pi \dfrac{l_2-2r}{l_1}}\Big)}{2\pi(l_1^2 - l_2^2)} \\
- \dfrac{l_2^3 \sin{\Big(\pi \dfrac{l_1-2r}{l_2}}\Big)}{2\pi(l_1^2 - l_2^2)}\Bigg] \text { if } \;\dfrac{|l_2 - l_1|}{2} \leq r\leq \dfrac{|l_1 +l_2|}{2}\\
k _ { S _ { 1 } \times S _ { 2 } } \left( r ; l _ { 1 } , l _ { 2 } \right) = 0 \quad \text { if } \quad r \geq \frac { l _ { 1 } + l _ { 2 } } { 2 }
\end{split}
\end{equation}
with $l_{mean} = (l_1 + l_2)/2$, $l_{min}=\min(l_1,l_2)$, and $l_{max}=\max(l_1,l_2)$.
The full covariance matrix can then be constructed by setting sparse covariance functions on all diagonal block elements and sparse-sparse cross-covariance functions on all non-diagonal block elements, e.g., for three cross-correlated covariance functions, the complete covariance matrix is given by  $K$ =
\begin{equation}
\begin{bmatrix}
k_{S_1} & w_{1,2} k_{S_1 \times S_2} & w_{1,3} k_{S_1 \times S_3} \\
w_{1,2} k_{S_2 \times S_1} & k_{S_2} & w_{2,3} k_{S_2 \times S_3} \\
w_{1,3} k_{S_3 \times S_1} & w_{2,3} k_{S_3 \times S_2} & k_{S_3} 
\end{bmatrix}
\end{equation}
with the cross-covariance amplitude terms, $w_{1,2}$, $w_{1,3}$, $w_{2,3}$, which can be determined by optimising the marginal log-likelihood (see equation \ref{eq_logl}) in conjunction with all other GP hyperparameters (amplitude, kernel length-scale). While these correlation terms between different geophysical properties can be also measured independently, the joint optimization has the advantage to find the best possible solution over the complete parameter space.

\subsubsection*{Joint Inversion}
Joint Inversion within a Bayesian framework corresponds to computing the posterior distribution of an ensemble of quantities of interest ${\phi_n}_{n=1}^N$ (e.g. $N$ distinct rock properties) at the locations of interest ${x_m}_{m=1}^M$ given $M$ sensor observations $\mathbf{y}$.
If we assume a GP prior over the functions $\phi$ and an isotropic likelihood model:
\begin{eqnarray}
\begin{aligned} \phi ( \mathbf { x } ) & \sim \mathcal { GP } \left( 0 , \mathcal{K} _ { \phi,\phi^{\prime}} \left( \mathbf { x } , \mathbf { x } ^ { \prime } \right) \right) \\ 
\mathbf { y } & = \mathbf { f } + \mathbf { \eta } \quad \text { with } \\ 
\eta & \sim \mathcal { N } ( \mathbf { \eta } | \mathbf { 0 } , \sigma_s ^ { 2 } \mathbf { I }, ) \end{aligned} 
\end{eqnarray}
where $\mathcal{K} _ { \phi,\phi^{\prime} }$ is the covariance function in the parameter space including cross-covariances between distinct quantities of interest ($\phi,\phi^{\prime}$),  and $\sigma_s^{2}$ is the noise variance of the sensor observations. 
This translates to a Gaussian prior over the quantities of interest:
\begin{equation}
\phi \sim \mathcal { N } ( \phi | \mathbf { 0 } , \mathbf { K} _ { \phi \phi^{\prime}  } ),
\end{equation}
where $\mathbf { K  _ { \phi \phi^{\prime}  } } $ is the (N$\times$N)$\times$(M$\times$M) covariance matrix obtained by evaluating the kernel $K_{\Phi,\Phi^{\prime}}(x,x^{\prime})$ at the locations $(x,x^{\prime})$ of interest and the covariance between distinct quantities $\Phi$ and $\Phi^{\prime}$. The covariance structure in the sensor space is computed as:
\begin{eqnarray}
 \mathbf { K } _ { y \phi } &=& \mathbf { G } \mathbf { K } _ { \phi \phi^{\prime} }\\\ 
\mathbf { K } _ { y y } &= &\mathbf { G } \mathbf { K } _ { \phi \phi ^{\prime}} \mathbf { G } ^ { T } + \sigma_s ^ { 2 } \mathbf { I }, 
\label{eq_regularisation}
 \end{eqnarray}
 where the joint forward model $\mathbf { G }$ is an operator matrix with diagonal block elements representing the distinct forward model operators and non-diagonal set to zero. For instance a joint forward model with three different forward models ($\mathbf {G_{1,2,3}}$) --- such as gravity, magnetic, and drill-cores --- is represented by the matrix
 \begin{equation}
\mathbf { G } = 
\begin{bmatrix}
[\mathbf{G_1}]_{j,m} & [\mathbf{0}]_{k,m} & [\mathbf{0}]_{l,m}\\
[\mathbf{0}]_{j,m}  & [\mathbf{G_2}]_{k,m} & [\mathbf{0}]_{l,m} \\
[\mathbf{0}]_{j,m} & [\mathbf{0}]_{k,m} & [\mathbf{G_3}]_{l,m}
\end{bmatrix},
\label{eq_Gmatrix}
\end{equation}
with M locations of interest and three sensor grids with J, K, and L number of sensors, respectively. Note that the uncertainty $\sigma$ of the sensor observations $\mathbf { y }$ is included in equation \ref{eq_regularisation} and can be either a constant $\sigma$ (same uncertainty for all sensors) or a vector that is given by the individual uncertainty of each sensor observation. In case the sensor uncertainty is unknown, it can be estimated in conjunction with the other GP hyperparameters by maximizing the marginal log-likelihood (see equation \ref{eq_logl}).

The predictive distribution for all quantities of interest $\Phi$ is given by
\begin{center}
\begin{eqnarray}
 \phi | \mathbf { G } , \mathbf { y } & \sim & \mathcal { N } ( \phi | \mathbf { \mu } _ { \phi | y } , \mathbf { \Sigma } _ { \phi | y } )  \; \; \text { with: }  \\ 
   \label{eq_inversion1}
  \mathbf{ \mu } _ { \phi | y }  & = & \mathbf { K } _ { y \phi } ^ { T } \mathbf { K } _ { y y} ^ { - 1 } \mathbf { y } \\ 
 \mathbf { \Sigma } _ { \phi | y }  & = & \mathbf { K } _ { \phi \phi } - \mathbf { K } _ { y \phi } ^ { T } \mathbf { K } _ { y y } ^ { - 1 } \mathbf { K } _ { y \phi },  
\label{eq_inversion2}
\end{eqnarray}
\end{center}
where $ \mathbf { \mu }$ is the mean at each location m and for each quantity n (matrix with M$\times$N elements), and  ${ \mathbf { \Sigma } _ { \phi | y } }$ the covariance matrix with (M$\times$N)$^2$ elements. Thus, different sensor types can be fused and processed jointly by taking advantage of cross-correlating the target quantities via a joint covariance function. This results in multivariate outputs and their respective uncertainties, which are important for subsequent decision-making processes as outlined in the next section. Another advantage of the Bayesian framework is the ability to calculate the marginal likelihood,  
\begin{equation}
\mathcal { L } ( \Theta ) = - \frac { 1 } { 2 } \mathbf { y } ^ { T } \mathbf { K } _ { y y} ^ { - 1 } \mathbf { y } - \frac { 1 } { 2 } \log \left| \mathbf { K } _ { y y} \right| - \frac { \log 2 \pi } { 2 } \sum _ { i = 1 } ^ { M } N _ { i }
\label{eq_logl}
\end{equation}
to optimize the parameters $\Theta$ of the GPs, which include the hyper-parameters of the covariance functions $k(x, x^{\prime})$, the noise variances $\sigma_s^2$, and optionally the sensor uncertainties and cross-covariance amplitude terms (for more details see section \emph{Hyper-Parameter Learning for GP Inversion}).
        
\subsection{Multi-Objective Bayesian Optimisation}  
Bayesian Optimisation (BO) is a powerful framework for finding the extrema of objective functions that are noisy, expensive to evaluate, or do not have a closed-form (e.g., black-box functions), or have no accessible derivatives. Bayes' rule is applied to derive the posterior estimate of the unknown function, given sample observations (e.g., sensor measurements) and prior knowledge. The prior is a GP model of the function, whose mean captures the estimated value of the function, and whose variance determines the level of uncertainty of the prediction in that particular location. 
The posterior is then used to query for the next most promising point based on objectives that are defined in an acquisition function $g$, which guides the search for a user-defined optimum. This is equivalent of finding the  $\mathbf { x } ^ { \star } = \arg \max _ { \mathbf { x } } g ( \mathbf { x } )$.  The BO algorithm maximizes this function that quantifies the benefit of choosing a specific location to be sampled. Once the maximum of $g$ is found and new measurements is taken at the proposed position, the model is updated with the new data, and the BO algorithm proposes a new point at each iteration step until the objectives are achieved or a maximum of iterations is reached.\par

\subsubsection*{Acquisition Functions}
The key of BO is the acquisition function $g$, which typically has to balance between (1) exploration, i.e., querying points that maximise the information gain and minimize the uncertainty of a model (e.g., of a geophysical model of a site), (2) exploitation, i.e., querying points that maximise the reward (e.g. concentrating search in the vicinity locations with high value such as minerals), and (3) minimize the number of samples given an expensive cost function for any new measurement (e.g., energy-cost of active sensors, or costly drill-cores). Several acquisition functions have been proposed \citep[for a practical overview, see, e.g.,][]{brochu2010tutorial} that make use of the mean and variance of the posterior, such as the probability of improvement, the expected improvement and the Upper/Lower Confidence Bound (UCB/LCB). The latter evaluates sample locations $\mathbf{x}$ in the form
\begin{equation}
ULCB (\mathbf { x }) = \mu ( \mathbf { x } ) \pm \kappa \cdot \sigma ( \mathbf { x } ),
\end{equation}
with $+(-)$ indicating the Upper (Lower) CB, the mean value for the prediction $ \mu ( \mathbf { x } ) $, the standard deviation $\sigma ( \mathbf { x } )$ of the GP posterior at $\mathbf{x}$, and the parameter $\kappa$ that balances the exploration-exploitation trade-off. \par

Finding the optimal $\kappa$ is not trivial and may depend on the user-specific objectives, or can be constraint by its cumulative regret in terms of maximal information gain \citep{srinivas2009}. If the parameter $\kappa $ is not dependent on any user-specific objectives, a simple estimate of $\kappa$ can be obtained under assumption of equal weight between the mean and the variance  :
\begin{equation}
\kappa =\dfrac{\sigma( \mathbf {\mu }( \mathbf { x } ))^2}{\sigma(\mathbf {\sigma} ( \mathbf { x } ))^2}. 
\end{equation}
Alternatively, it is possible to optimize the UCB acquisition function over a uniform range of $\kappa$ values and to select the most frequent solution. Moreover, various extensions to the acquisition function have been proposed to accommodate the increasing complexity of remote sensing applications \citep[ e.g., environmental monitoring,][]{marchant2012}.

\subsubsection*{Multiple Objectives: Pareto Front and Expected Hyper-volume Improvement}
Typically, multiple objectives are conflicting, and no solution may exists where all objectives can be minimized at once. In this case the task is then to find the set of optimal compromise solutions, where at least one objective improves without reducing any of the other objectives. The optimal set of these solutions is called Pareto set, named after Vilfredo Pareto, who applied this concept to study economic efficiencies \citep{cirillo2012}. Graphically it can be described as a frontier of optimal points in a hyper-dimensional space, where each dimension corresponds to one objective. The frontier defines all points of a sample that dominate over others, so that it is not possible to improve an objective without degrading the others.\par

\begin{figure}[!t]
\centering
\includegraphics[width=4.5in]{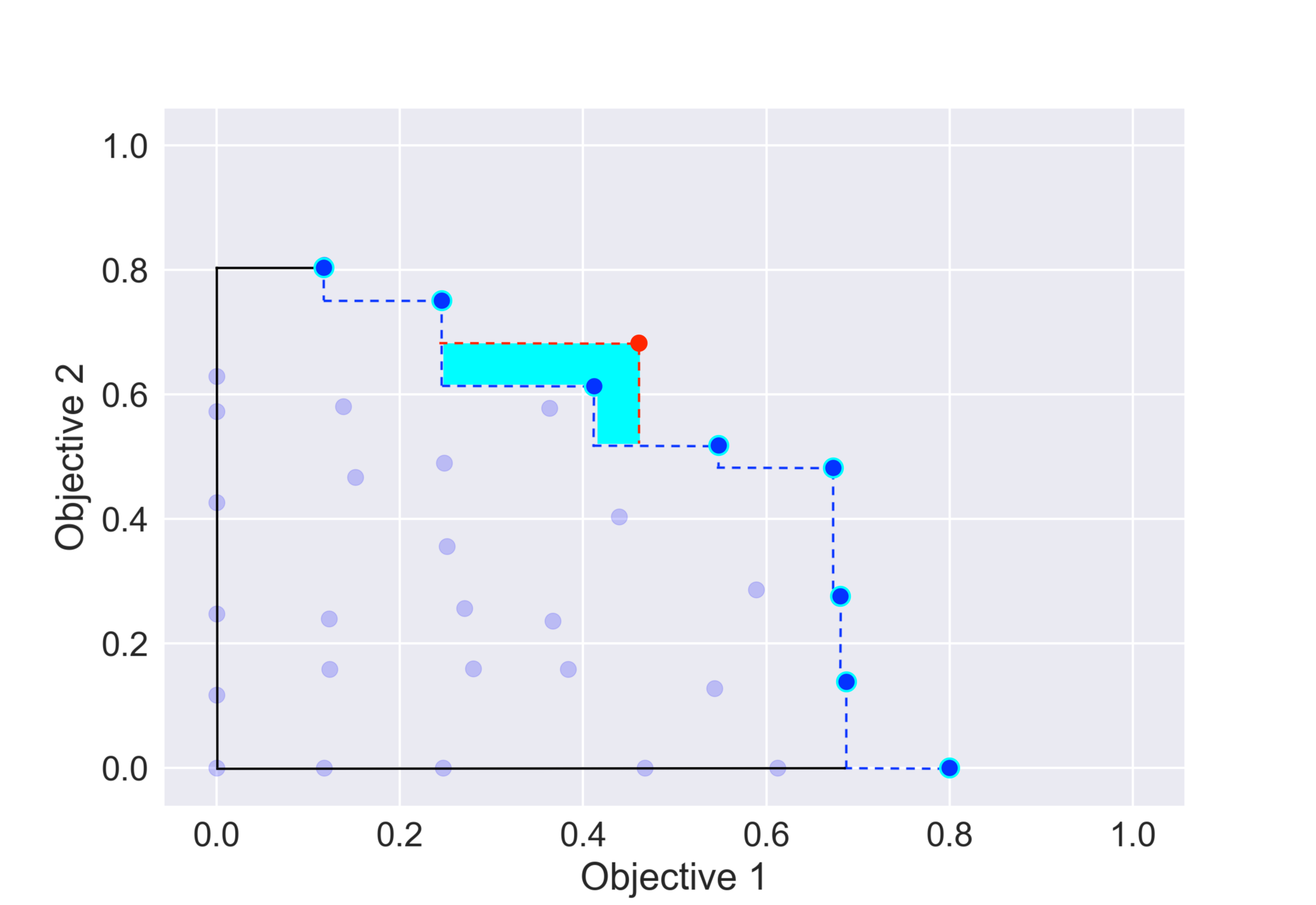}
\caption{Illustration of a Pareto Front (dashed blue line) with two objective functions to be maximized (e.g., mineral gain vs drillcore costs). The Expected Improvement corresponds to the additional area (filled blue) spanned by the new Pareto Front (dashed red line) given by a new sampling point.}
\label{fig_paretofront}
\end{figure}

To query a new sample point, we calculate the Expected Hyper-volume Improvement \citep[EHVI,][]{emmerich2005} with regards to the current optimal Pareto set. The Pareto Front and EHVI is illustrated in Figure~\ref{fig_paretofront} for two objectives, which reduces to the problem to query the sample point that maximizes the \textit{Area of expected improvement} that is spanned by the new Pareto Front. This is straightforward if the number of evaluation points is not too high, otherwise computationally efficient grid-search or variational methods can be applied to speed-up the computation.  For larger number of  objectives $N_{Obj}$ and evaluation points, the integration of the EHVI becomes computationally very expensive as it increases exponentially with $N_{Obj}^3$ and several decomposition methods have been proposed \citep{emmerich2008,bader2011,feliot2017} by  partitioning the integration domain into hyper-rectangles or using sequential Monte Carlo techniques. \par

Multi-objective optimisation is particularly important for joint inverse problems, which generate posteriors via a GP prior with multiple outputs such as different geophysical properties. To find the optimal new sampling point, we calculate the EHVI given $N$ objectives. In this paper each objective function $f_i$ is defined by maximizing the UCB
\begin{equation}
UCB_i (\mathbf { x }) = \mu_i ( \mathbf { x } ) + \kappa_i \cdot \sigma_i (\mathbf { x }) - \gamma_i c(x)
\label{eq_ucb}
\end{equation}
with the mean value for the prediction $ \mu_i (\mathbf{x}) $, the variance $\sigma_i^2 ( \mathbf{x} )$, and a cost function  $c(x)$, which is defined by the cost of obtaining a measurement at the sample point $x$. The parameter $\kappa_i$ and $\gamma_i$ define the trade-off in exploration-to-exploitation and gain-to-cost, respectively. For example, maximizing the mean value can be beneficial if the goal is to sample new data at locations with high density or mineral content, and not only where the uncertainty is high. At each iteration of the BO, new potential sample points are evaluated given their EHVI. The BO algorithm stops if none of the proposed new sample points improves any of the objectives without reducing any of the other.

\section{Experiments with Simulated Data} \label{sec:simulated}

\subsection{Geophysical Forward Models and Sensor Data Characterisation} 
	
To test our proposed BO algorithm, we first need to define the set of forward models that transform the localised measurement of a sensor grid into a 3D representation of geophysical properties of a region. Inversion can be solved for linear forward models which  follow the relation $\mathrm { y } = \mathrm { G } \phi$ where $\mathrm { y }$ is the vector of  observations; $\phi$ is the vector of unknown parameters at the 3D locations; $\mathrm {G }$ is a known $N \times M$ sensitivity matrix that relates the values of the geophysical property at different locations to the observations. The most common geophysical linear forward models are gravity and magnetic forward models as described in the following.\par 

\subsubsection*{Gravity Forward Model}

The gravity forward model describes the relation between the earth's density distribution and the gravitational force measured by sensors above the surface by simulating Newton's law. The gravity forward model is defined by using Li's tractable approximation for a 3D field of constant density prisms \citep{li1998} and can be determined analytically \citep{nagy2000gravitational}. A detailed form of the sensitivity matrix calculation is given in Appendix A.

\subsubsection*{Magnetic Forward Model}
Magnetic susceptibility depends on the Earth's magnetic mineral content such as iron-rich minerals including magnetite and haematite below the surface and is measured by the response of their magnetic dipoles induced by the Earth's magnetic field.  The induced magnetic field calculation uses Li's tractable approximation for a 3D field of prisms of constant susceptibility (neglecting residual permanent magnetism), and Earth's local magnetic field. The joint GP inversion takes into account a covariance that exists between density and magnetic susceptibility, which does provide additional insight into the region's geophysical properties. For instance, magnetite has a high magnetic susceptibility and density, while haematite has a similar density to magnetite, but a much lower magnetic susceptibility.

\subsection{BO sampling} 
We define the goal of the BO experiment to find the optimal location where to place a new sample measurement as given by a set of acquisition functions. The experiments are carried out in the following steps:
\begin{enumerate}
\item Generate a simulated 3D voxel cube with geophysical structures given by their density and magnetic susceptibility (described by the vector $\phi$).
\item Calculate response $\mathbf{y}$ of gravity and magnetic sensors at the surface, and of pre-existing drill-core samples.
\item Solve joint GP inversion to calculate the posterior distribution of geophysical properties in 3D as defined by their mean and variance for each voxel (equations \ref{eq_inversion1} and \ref{eq_inversion2}). 
\item Find the $x,y$ location for new drill-core that maximizes the BO acquisition function(s) (e.g., given by UCB, equation \ref{eq_ucb} or Pareto optimisation). 
\item Add selected drill-core data to measurements and repeat Steps 2 to 4 until the maximum of iterations (here 25) or final objective is achieved.
\end{enumerate}
While more drill-core parameters can be added (drill length, dip-, and azimuth-angle), we assume for simplicity in this study vertical drill-holes with constant length equal to the cube height.

\subsection{Synthetic Model Setup} 

\begin{figure}[!t]
\centering
\includegraphics[width=2.9in]{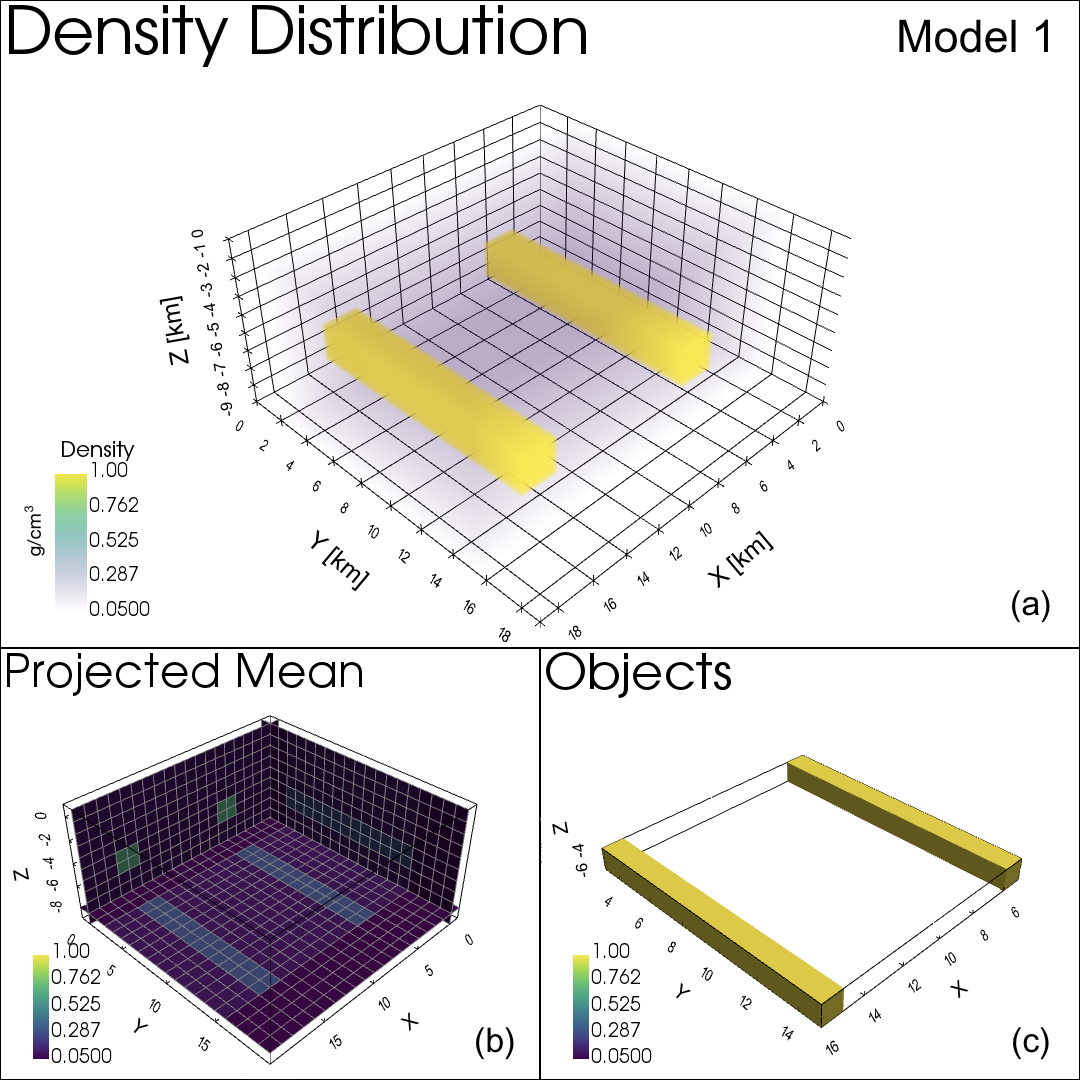}
\includegraphics[width=2.9in]{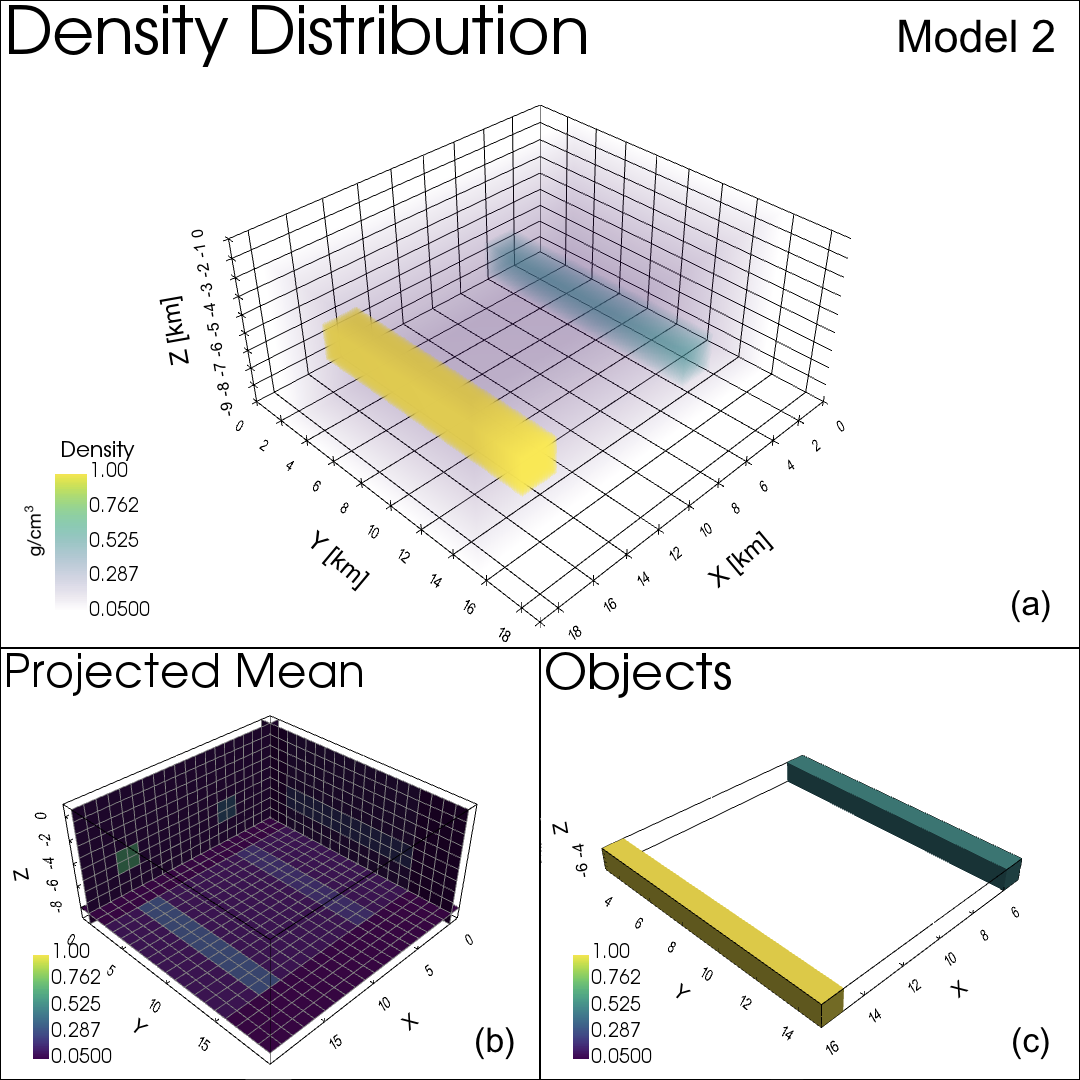}
\includegraphics[width=2.9in]{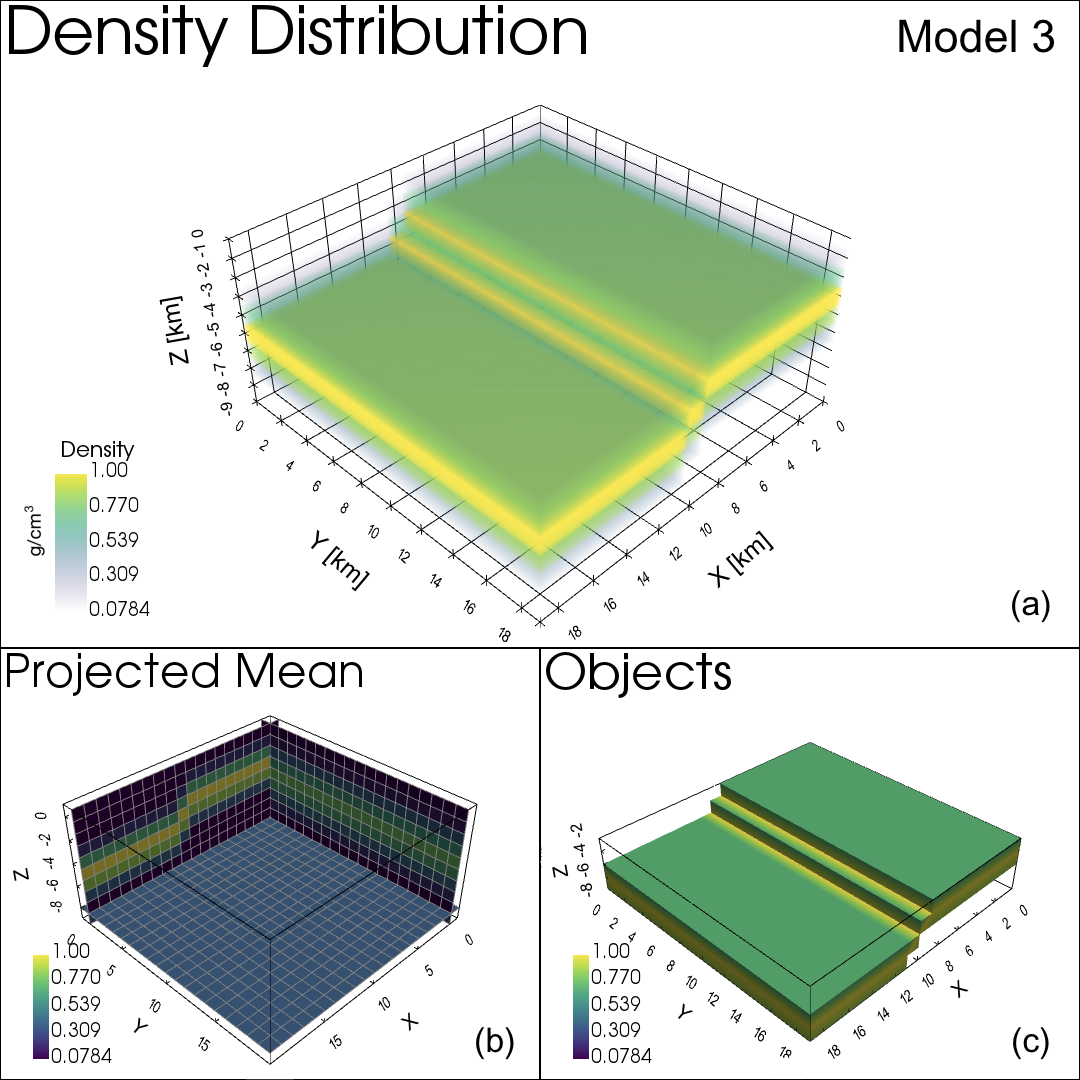}
\includegraphics[width=2.9in]{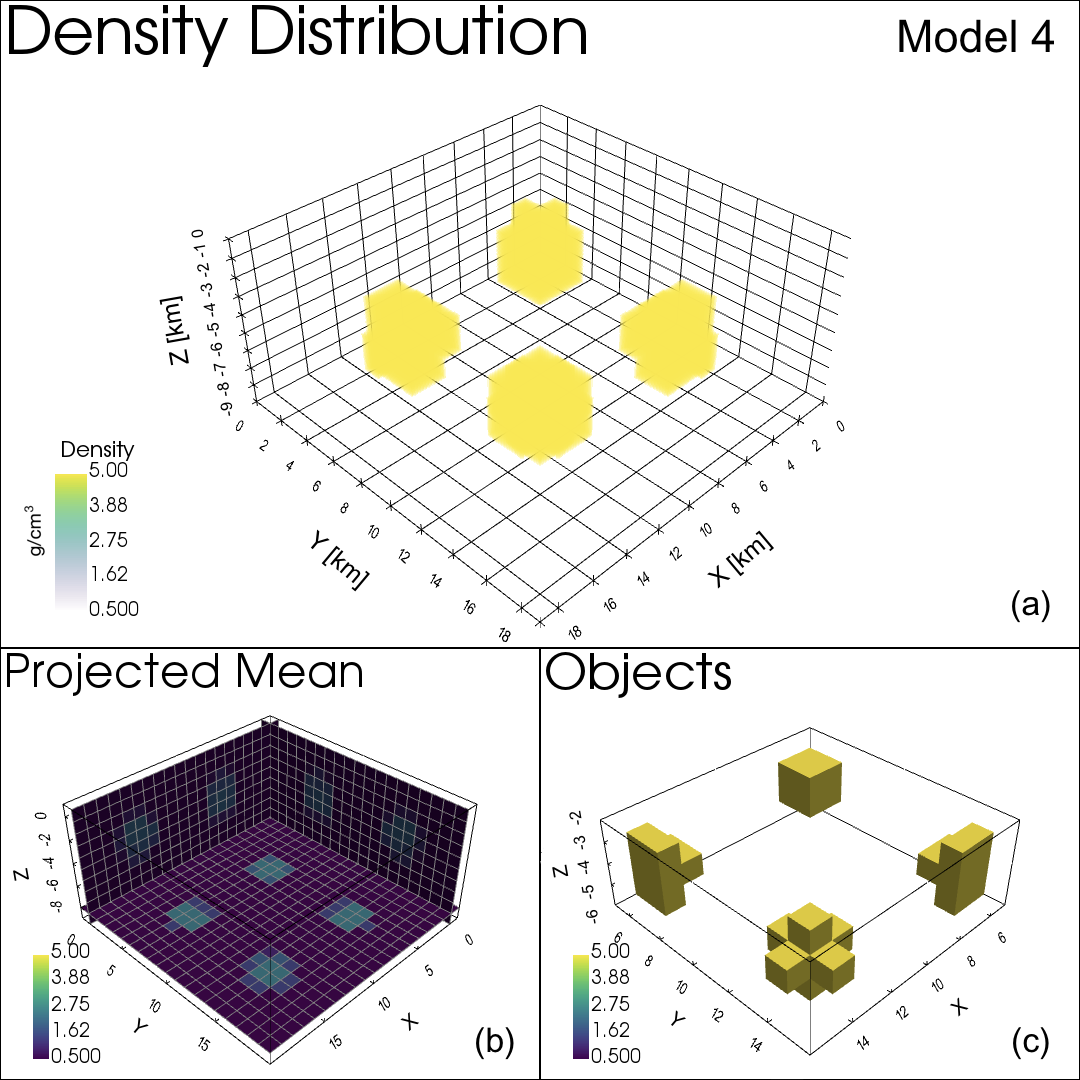}
\caption{The 3D synthetic models' density distribution for the four different models: two-even cylindrical body model (Model 1),  two-uneven cylindrical body model (Model 2), three-layer model (Model 3), and clumpy model (Model 4). The magnetic susceptibility follows a similar distribution with varying correlation coefficients (see text). Each of the four models shows (a) the volumetric 3D density across an area of of $20\times20$km with 10km depth, (b) the mean projected density, and (c) the core model object bodies.}
\label{fig_modeloverview}
\end{figure}

To test the joint GP inversion and the proposed BO algorithm, a set of four different synthetic geophysical models are created as shown in Figure~\ref{fig_modeloverview}:
\begin{enumerate}
\item Even Cylinder Model: Two even cylinder dipping bodies, which consist of slabs of dense, magnetically susceptible material, and are surrounded by a uniform low density and magnetic medium.
\item Uneven Cylinder Model: Similar to the two even cylinder dipping bodies as described above, but with density ratio 2:1 and different density-to-magnetic susceptibility correlation coefficients.
\item Layer model: Multiple-layer folded structure, with the central layer consisting of dense, magnetically susceptible material surrounded by two layers of material with lower density and magnetic susceptibility. 
\item Clumpy model: Four spherical high density and magnetic bodies, at different depths and with different density-to-magnetic susceptibility correlation coefficients.
\end{enumerate}
All of the simulated environments include three sensor types: (1) 2D surface gravity measurements, (2) 2D surface magnetic sensors, and (3) vertical drill-cores for direct local measurements of rock density.

\subsubsection*{Geometry and Voxelisation}
The region's geophysical properties  are reconstructed in a cube with an extent of 20~km $\times$ 20~km in horizontal and 10~km in vertical direction. The gravitational and magnetic sensors are distributed over a 20 $\times$ 20 uniform distributed sensor grid. The final reconstructed scene is represented by a voxel grid of size 20 $\times$ 20 (horizontal) $\times$ 10 (vertical), resulting in total of 4000 voxels. For visual simplification the voxel centre positions $x,y$ are aligned with the positions of the surface sensor grid and we assume that the sensor grid is on a flat surface at constant surface height. 

\subsection{Hyper-Parameter Learning for GP Inversion}
As given by equation \ref{eq_logl}, the GP inversion framework provides a calculation of the marginal likelihood, which allows us to optimize the GP hyperparameters, i.e., the length-scale $l$ and the noise variance $\sigma_s^2$ as regularisation term. As optimisation strategy we apply a global optimisation (e.g., NLopt) as starting point for subsequent local optimisation \citep[e.g., COBYLA][]{powell1994direct} to obtain an optimum to a greater accuracy. We assume the same hyperparameter length-scale for the entire cube (see Discussion for including stationary dependence of hyperparameters).  Using the synthetic test bodies as described in the previous section, we have tested a variety of GP non-stationary kernels, including the squared exponential, Matern32, as well as sparse kernels. The selection of the optimal kernel and GP hyperparameters is based on their marginal likelihood and the comparison between reconstructed and true geophysical properties. Overall, the sparse covariance kernel performed best given the most robust solution over a wide range of synthetic bodies, with an optimal GP length-scale $l$ of five voxel lengths (equal one quarter of the region's length) and $\sigma_s^2$ of half the standard deviation of the normalised sensor measurements. The correlation amplitude $w_{D-M}$ between density and magnetic susceptibility depends on the geochemistry of the rock type, and we have included in our simulation a range of correlation coefficients from 0.0 (no correlation at all) to 1.0 (complete correlation).\par

\subsection{Results on Dipping Body Model}
\begin{figure}[!t]
\centering
\includegraphics[width=2.8in]{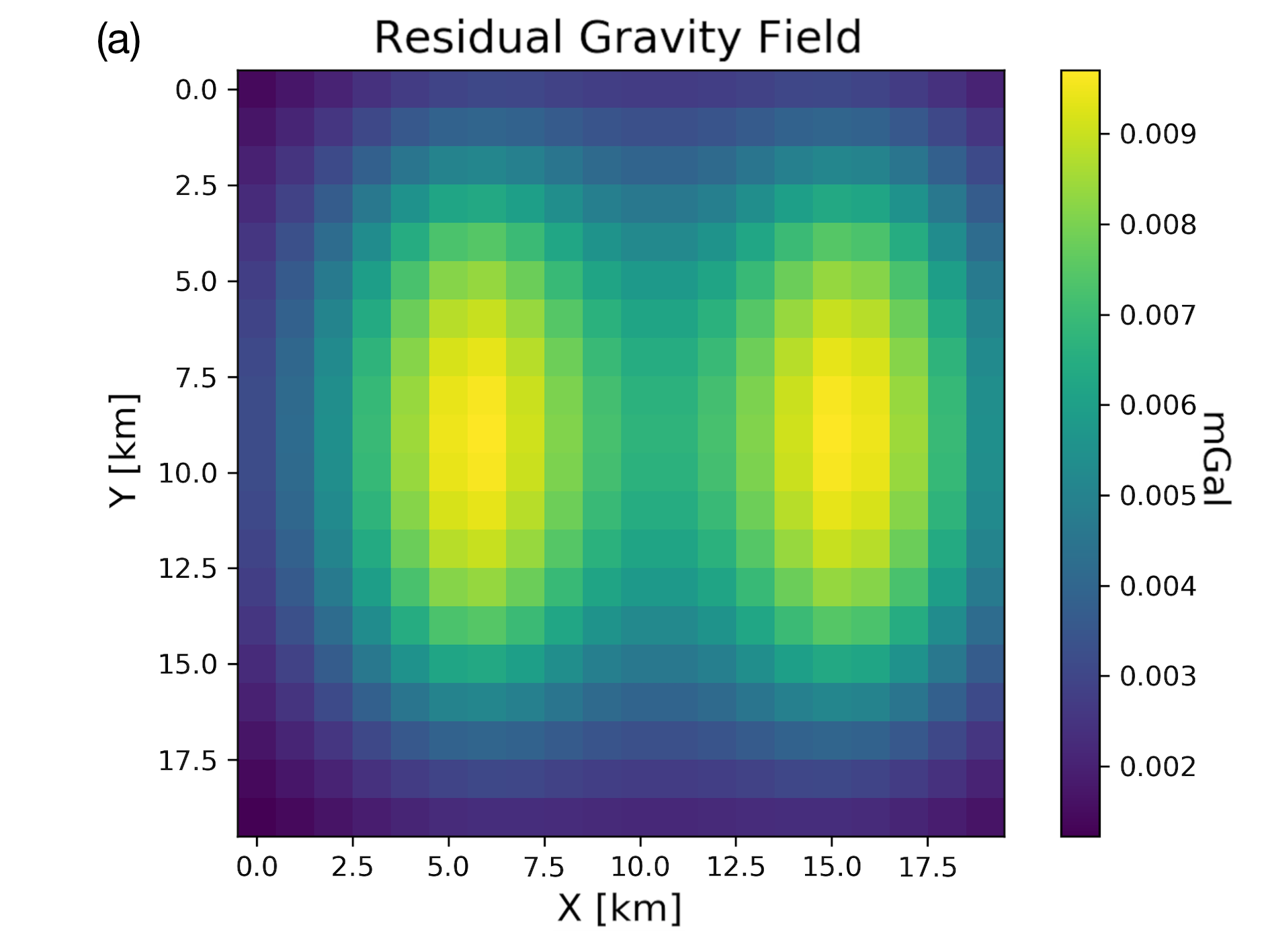}
\includegraphics[width=2.8in]{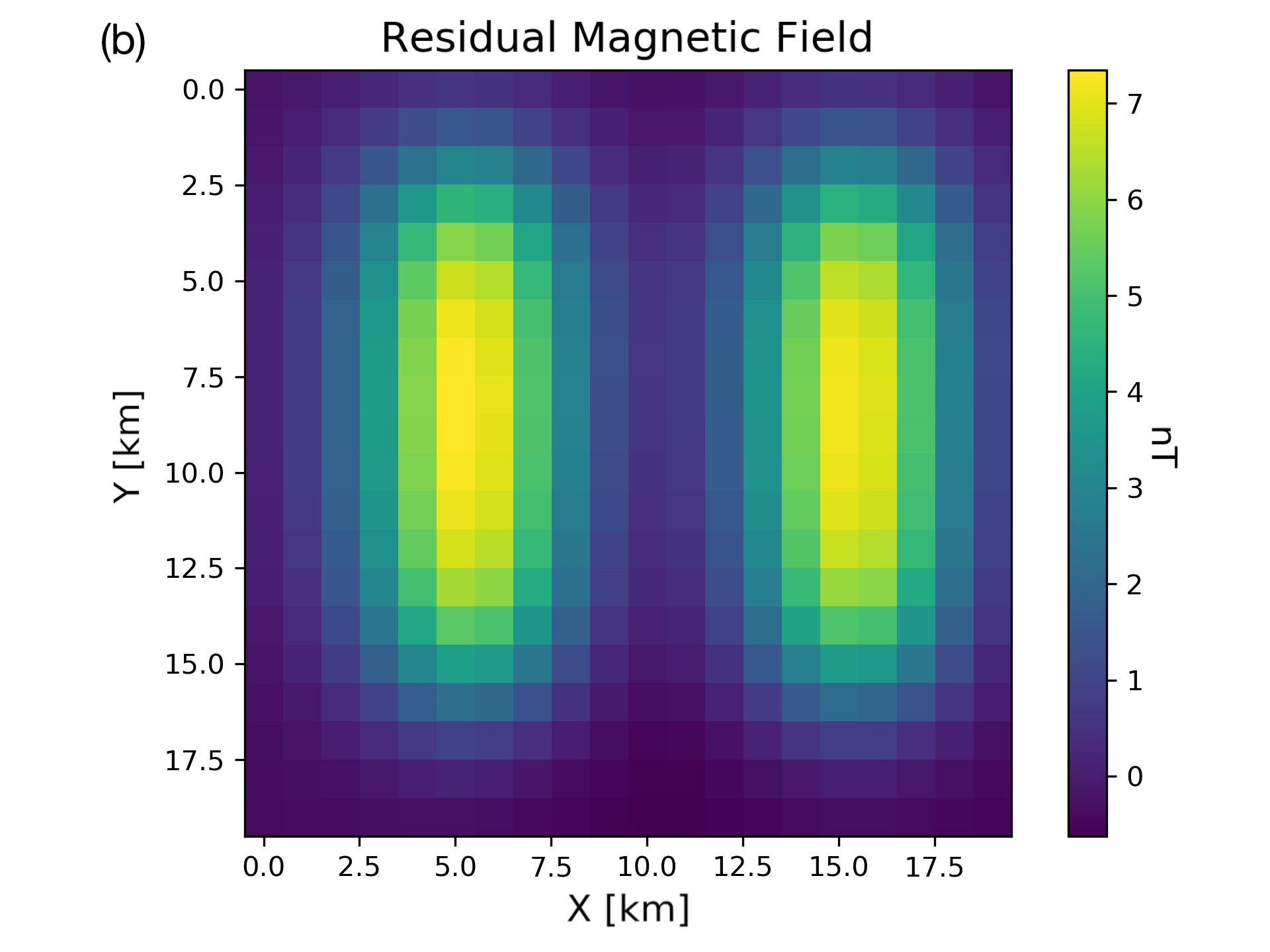} \\
\includegraphics[width=2.8in]{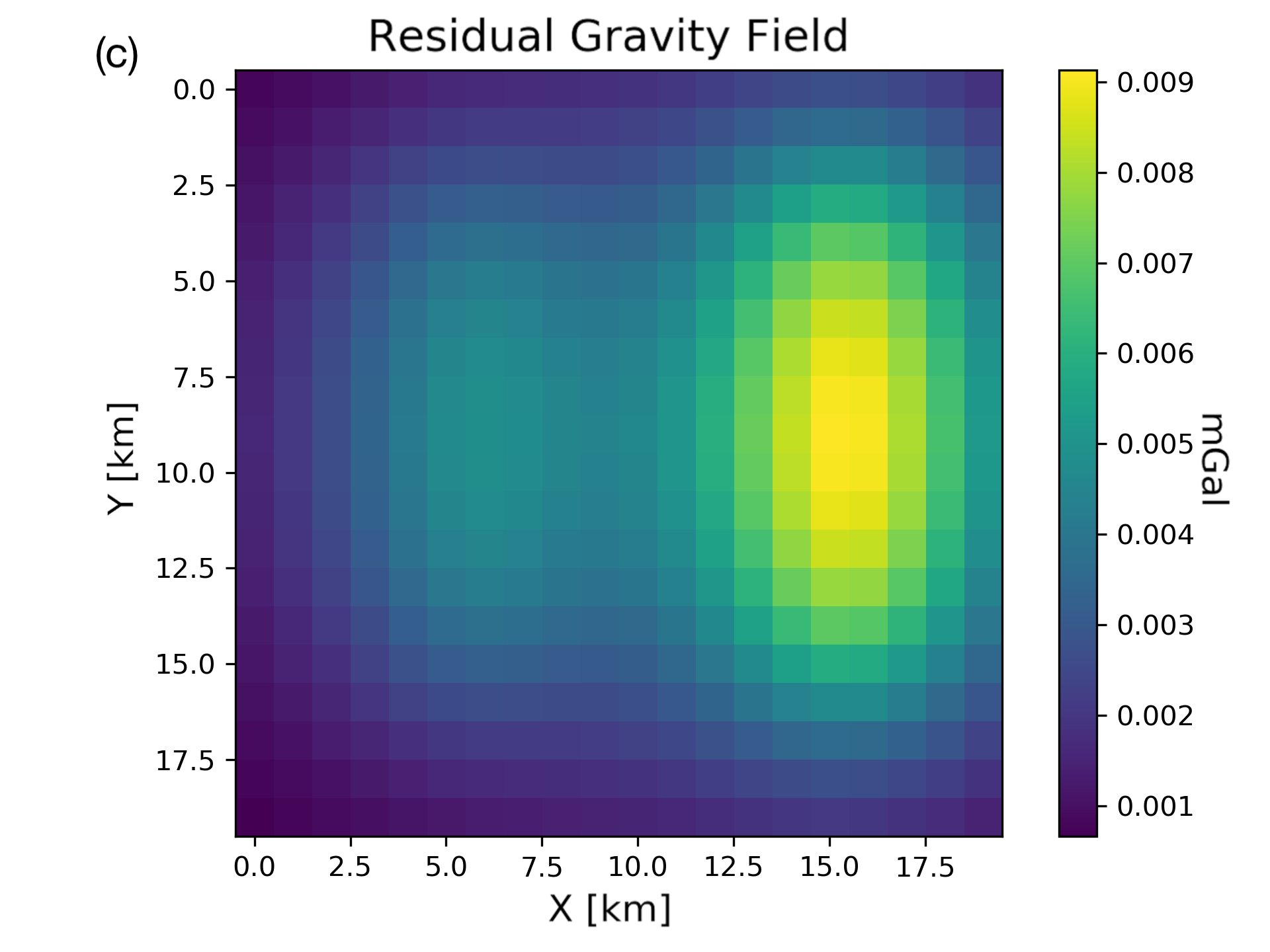}
\includegraphics[width=2.8in]{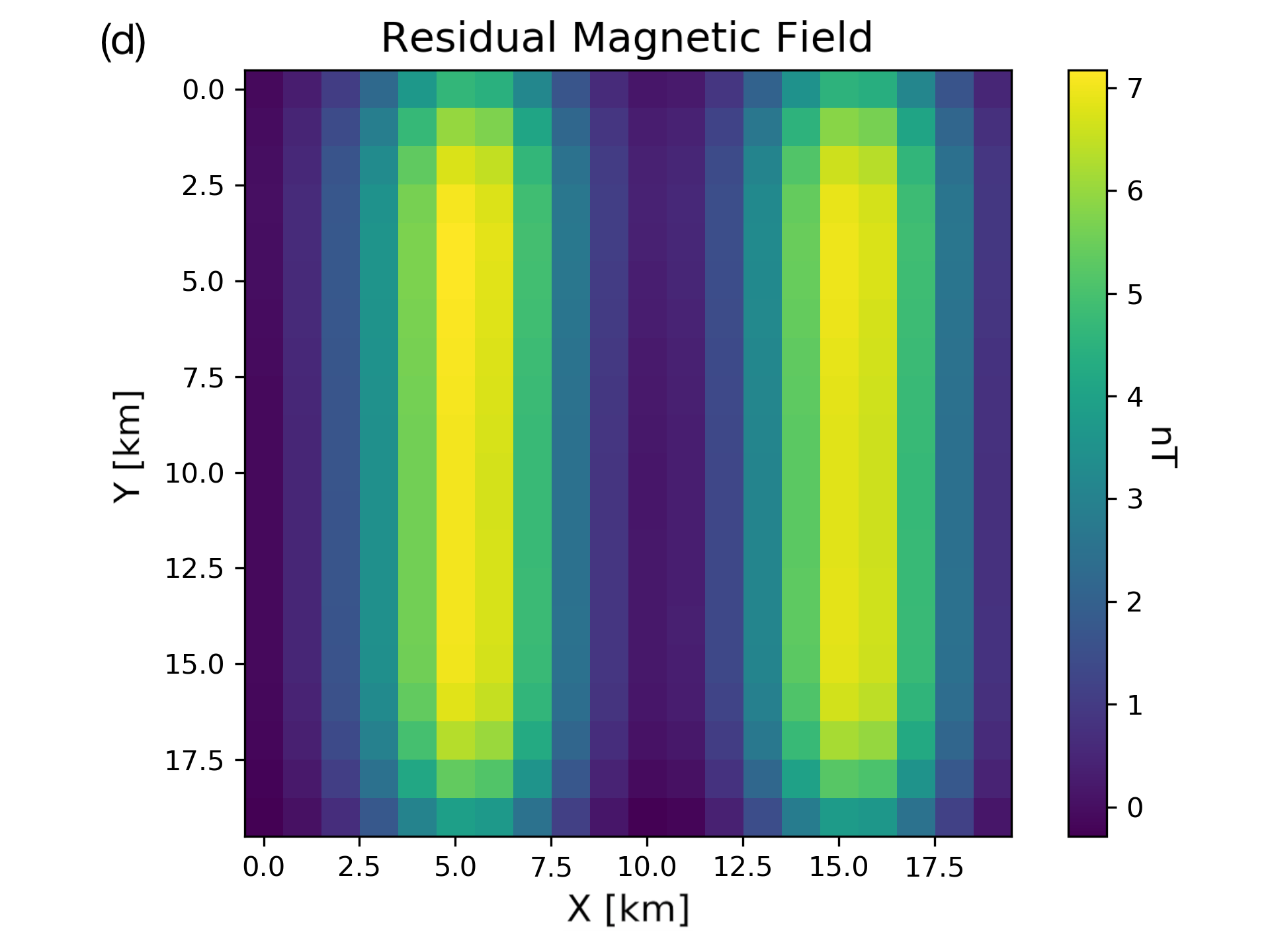}
\caption{(a) The gravitational anomaly field and (b) magnetic anomaly field on the surface as measured by sensors for scenario 1. (c) The  gravitational anomaly field and (d) magnetic anomaly field for scenario 2, see text.}
\label{fig_2cyl_gravmagn}
\end{figure}

\begin{figure}[!t]
\centering
\includegraphics[width=2.8in]{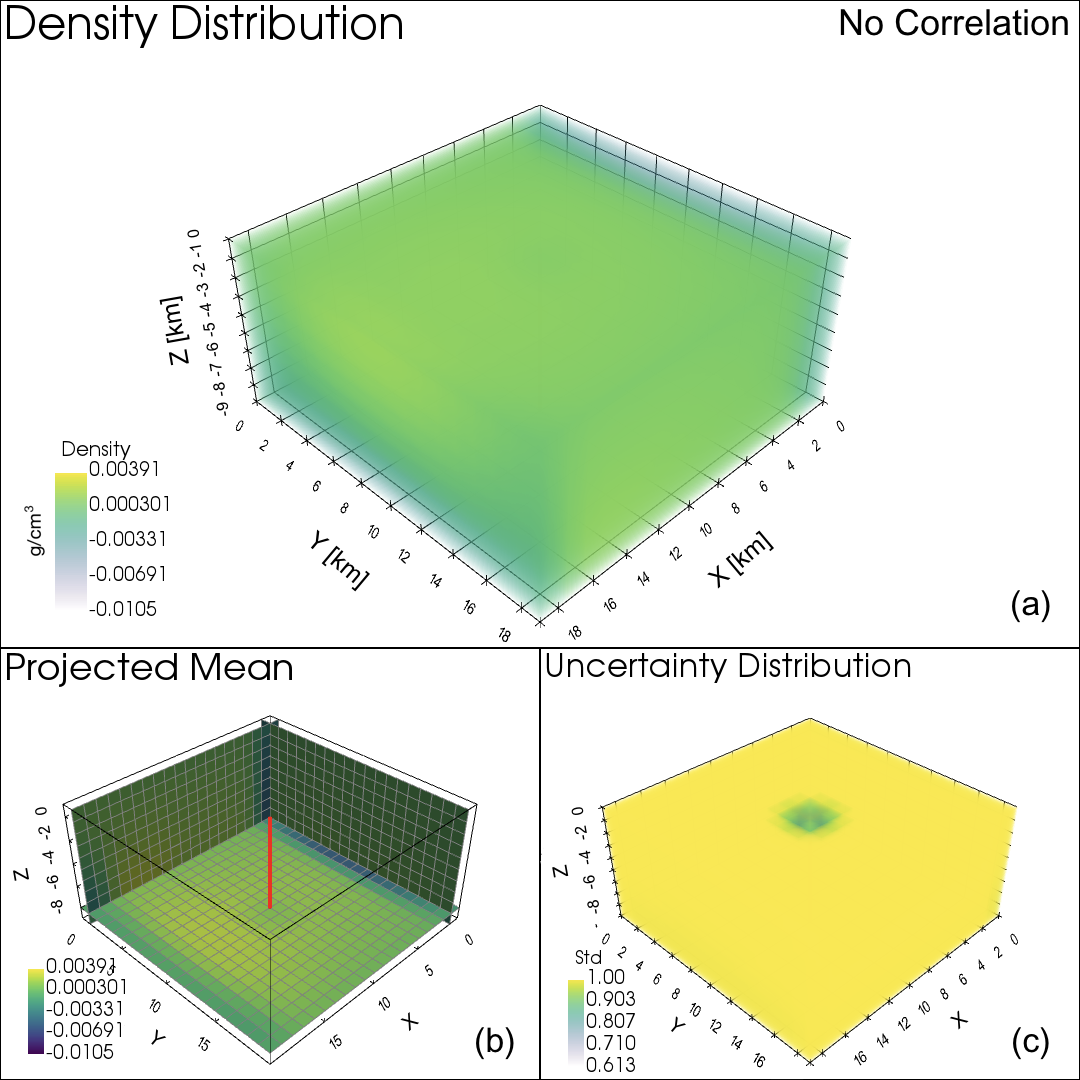}
\includegraphics[width=2.8in]{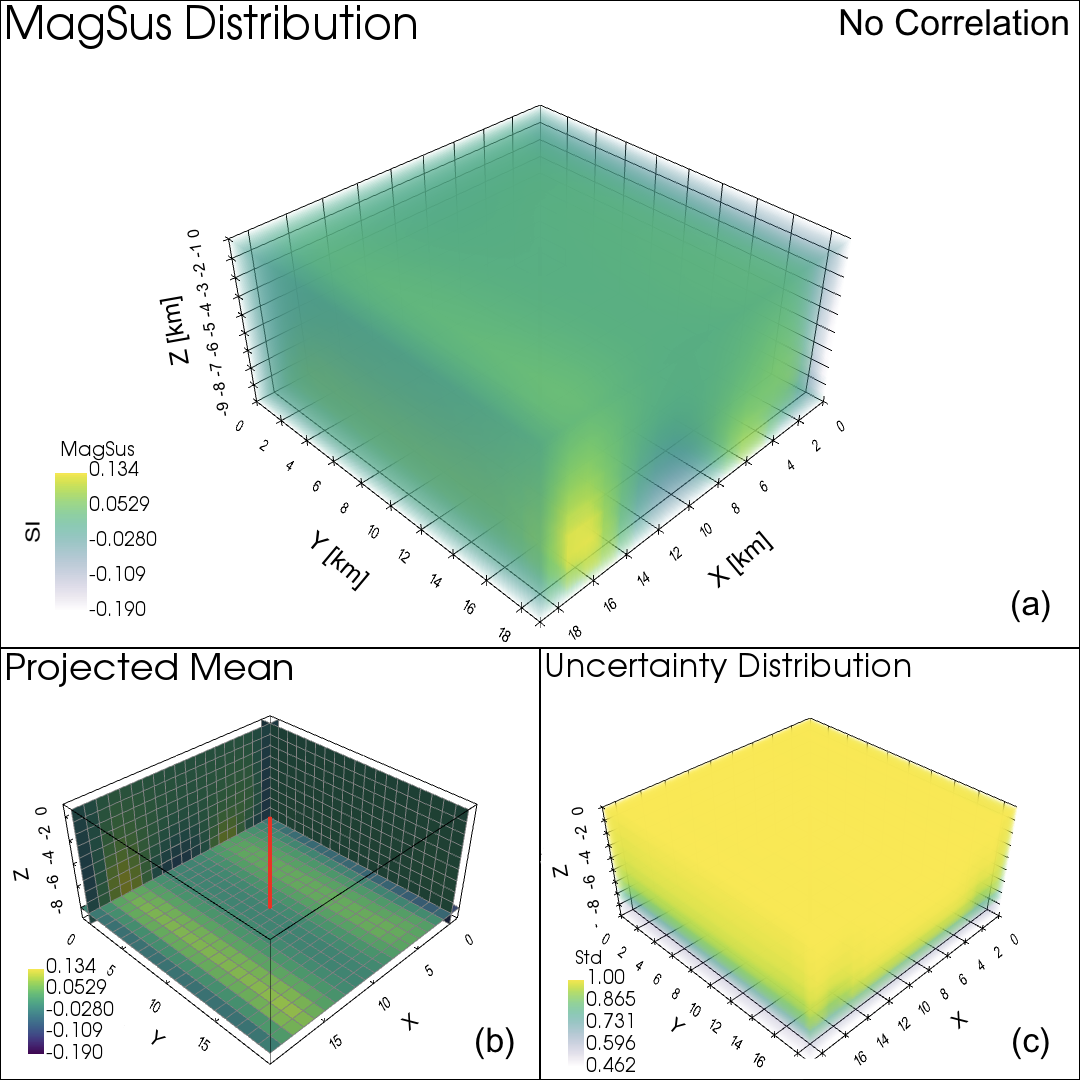}
\includegraphics[width=2.8in]{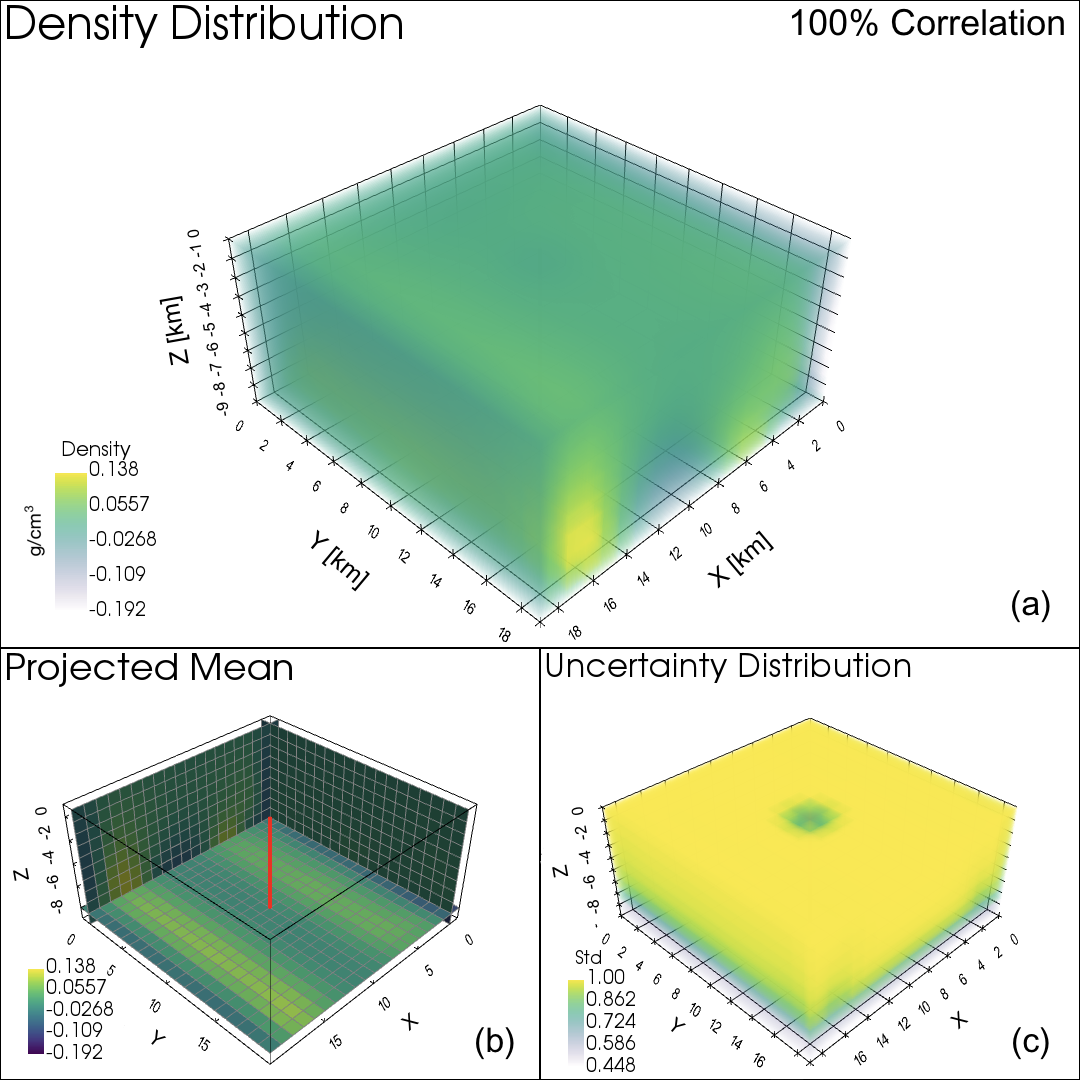}
\includegraphics[width=2.8in]{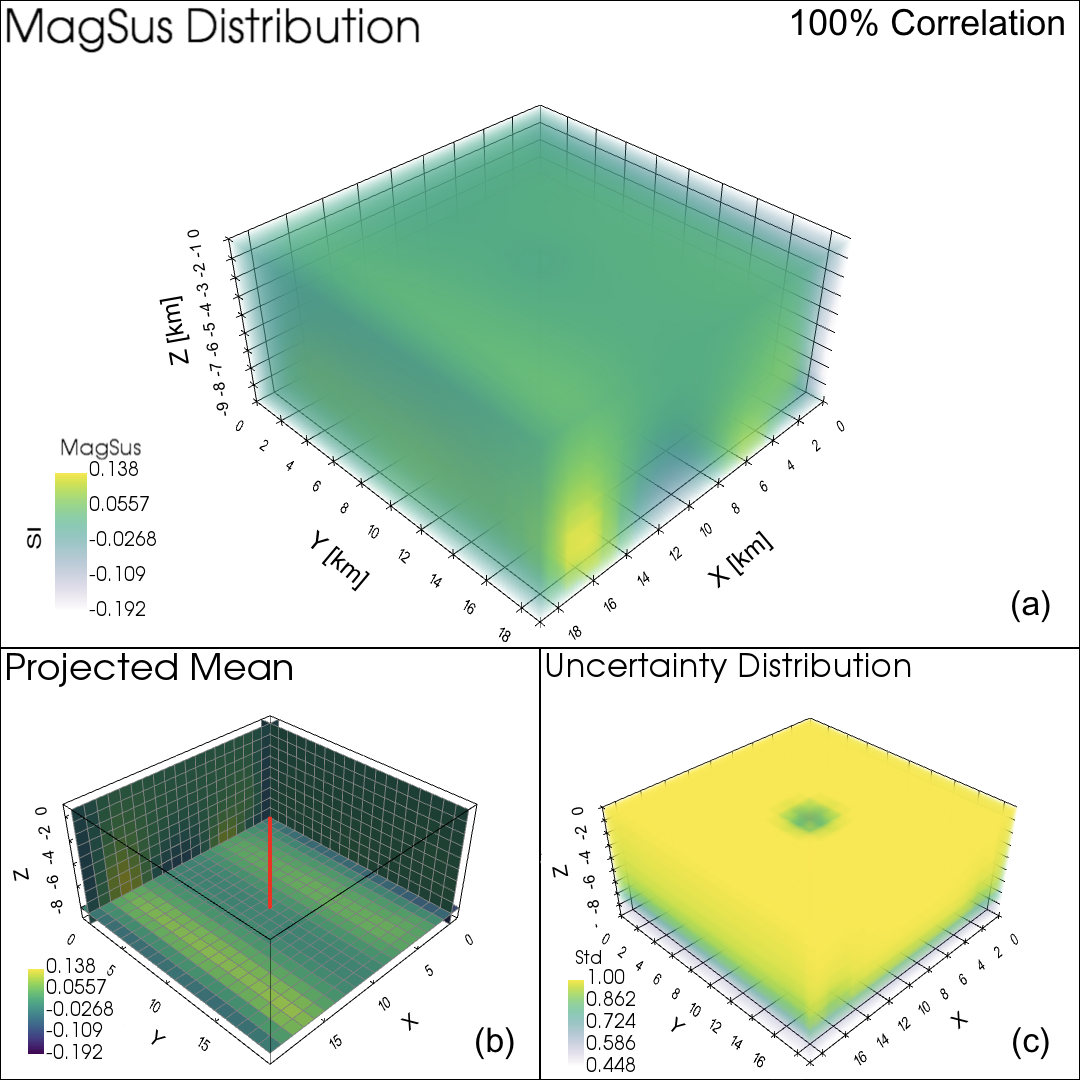}
\caption{The reconstructed density (Density Distribution) and magnetic susceptibility model (MagSus Distribution) based on only gravity, magnetic sensors and one central drill-core (red line), shown without correlation between density and magnetic susceptibility (No correlation, $w_{D-M} = 0$) and with $w_{D-M} = 1$ (100\% correlation) between density and magnetic susceptibility. Sub-panels (a,b,c) are defined as in Figure 3.}
\label{fig_2cyl_3Drec0}
\end{figure}

The synthetic model consists of two dipping bodies which are two separate, horizontally aligned cylinders of dense, magnetically susceptible material as shown in Figure~\ref{fig_modeloverview}. The surrounding medium is uniform with low density and magnetic susceptibility. In the first scenario the two dipping bodies have the same mass and same rock chemistry with $w_{D-M} = 1$. The second scenario consists of two dipping bodies with different mass and distinct rock chemistry with $w_{D-M} = 1$ for one cylinder and $w_{D-M} = 0.5$ for the other. The gravity and magnetic sensitivities are calculated from their forward models and are shown in Figure~\ref{fig_2cyl_gravmagn}.
An example solution of the joint inversion is shown in Figure~\ref{fig_2cyl_3Drec0} given only one central vertical drill-hole.\par

\begin{figure}[!t]
\centering
\includegraphics[width=2.8in]{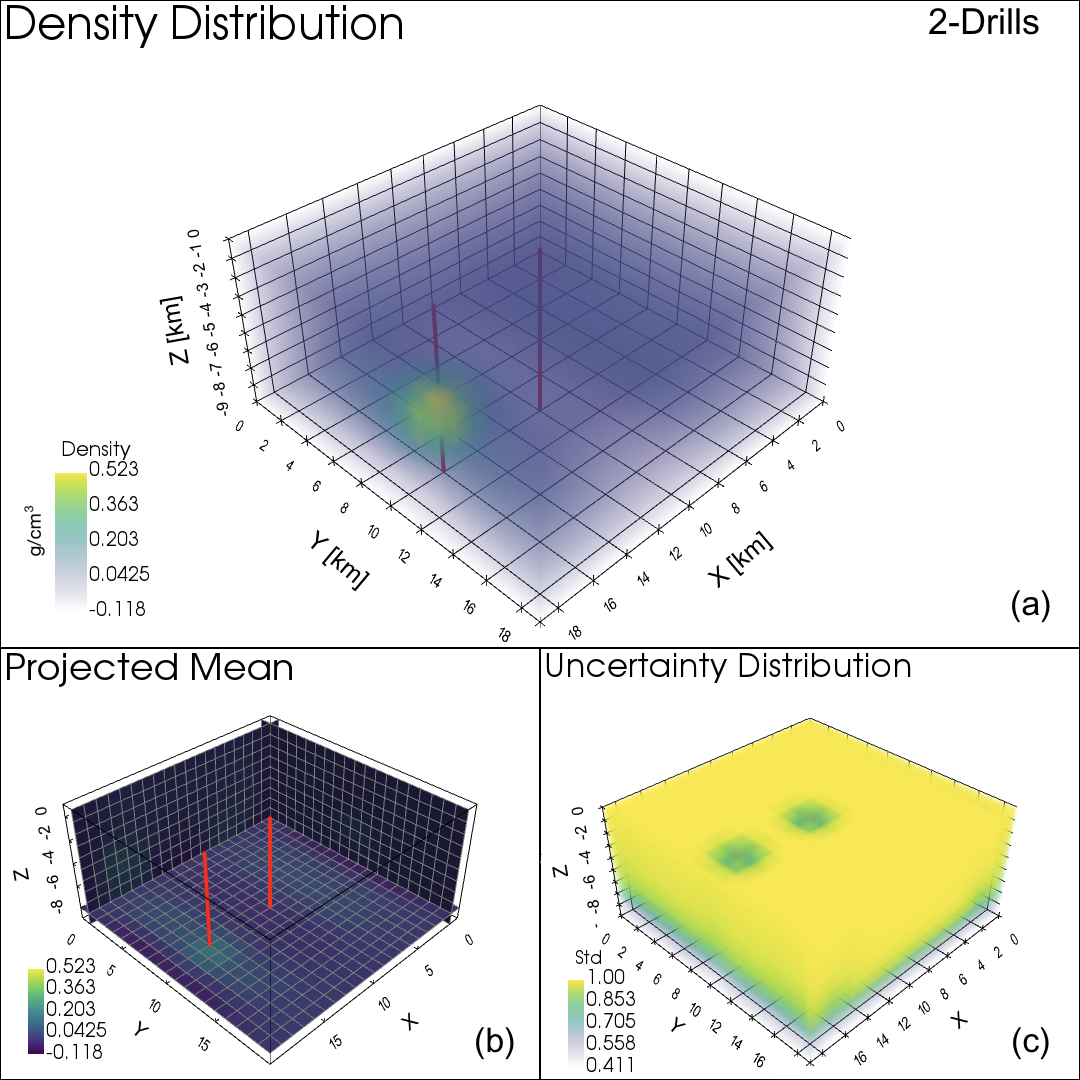}
\includegraphics[width=2.8in]{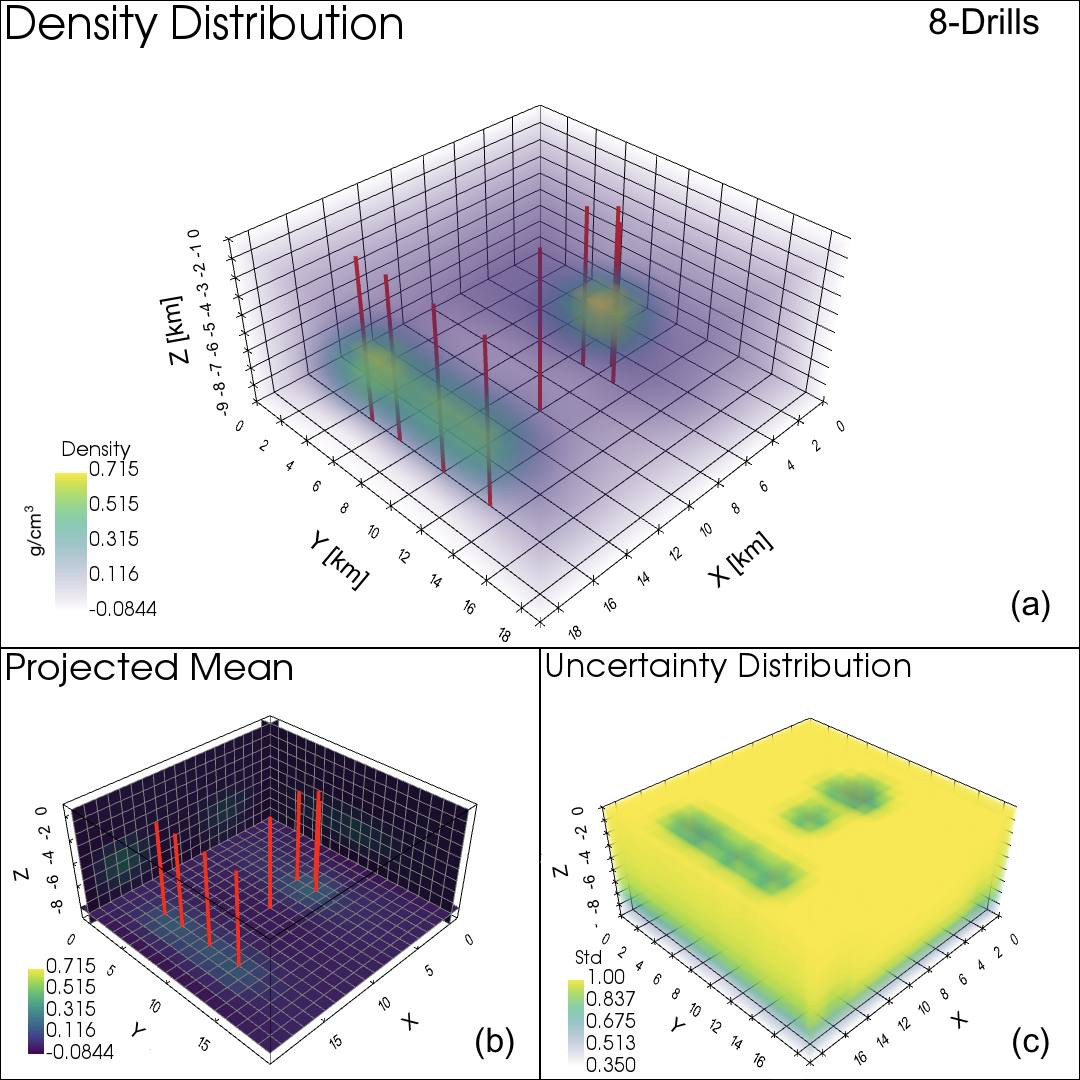}
\includegraphics[width=2.8in]{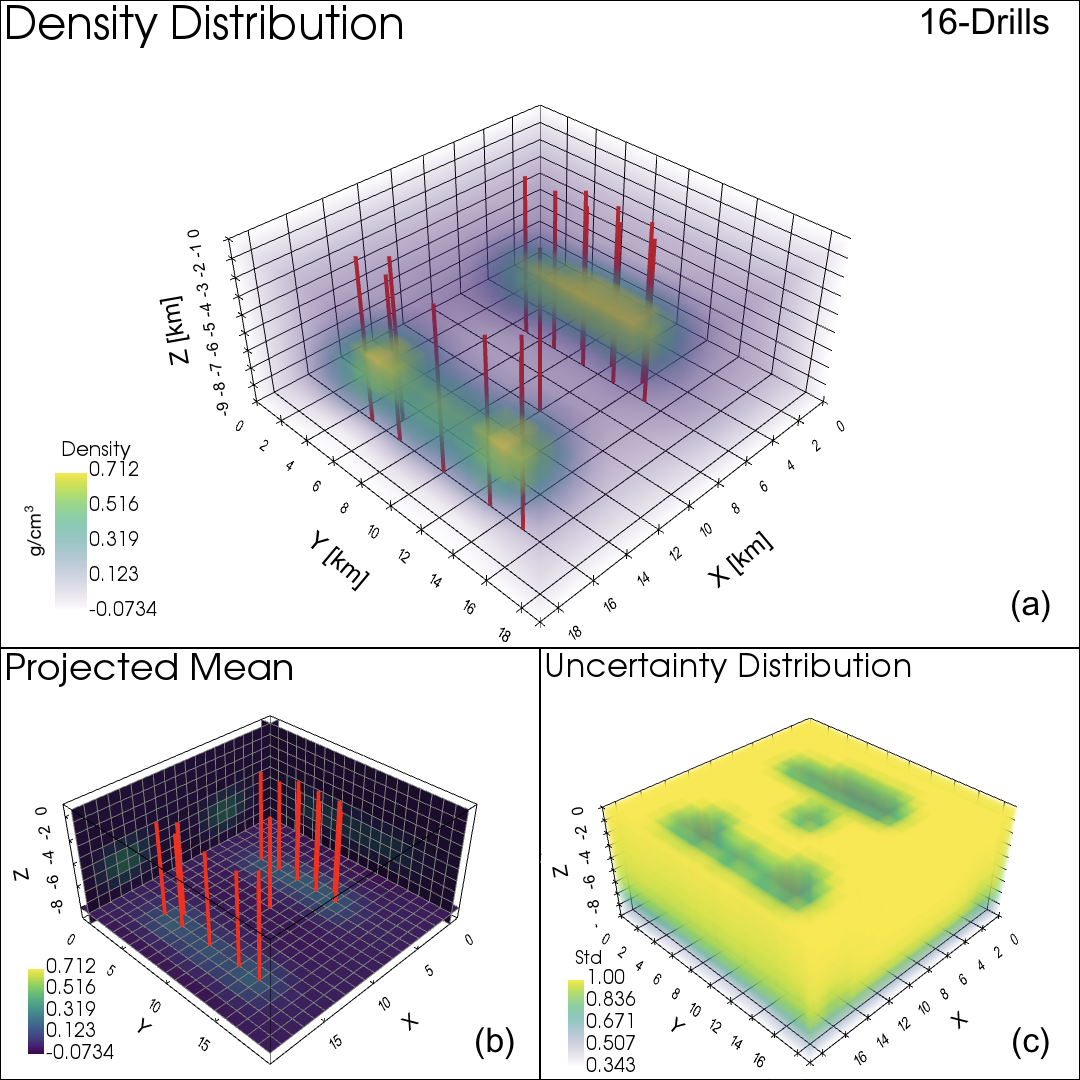}
\includegraphics[width=2.8in]{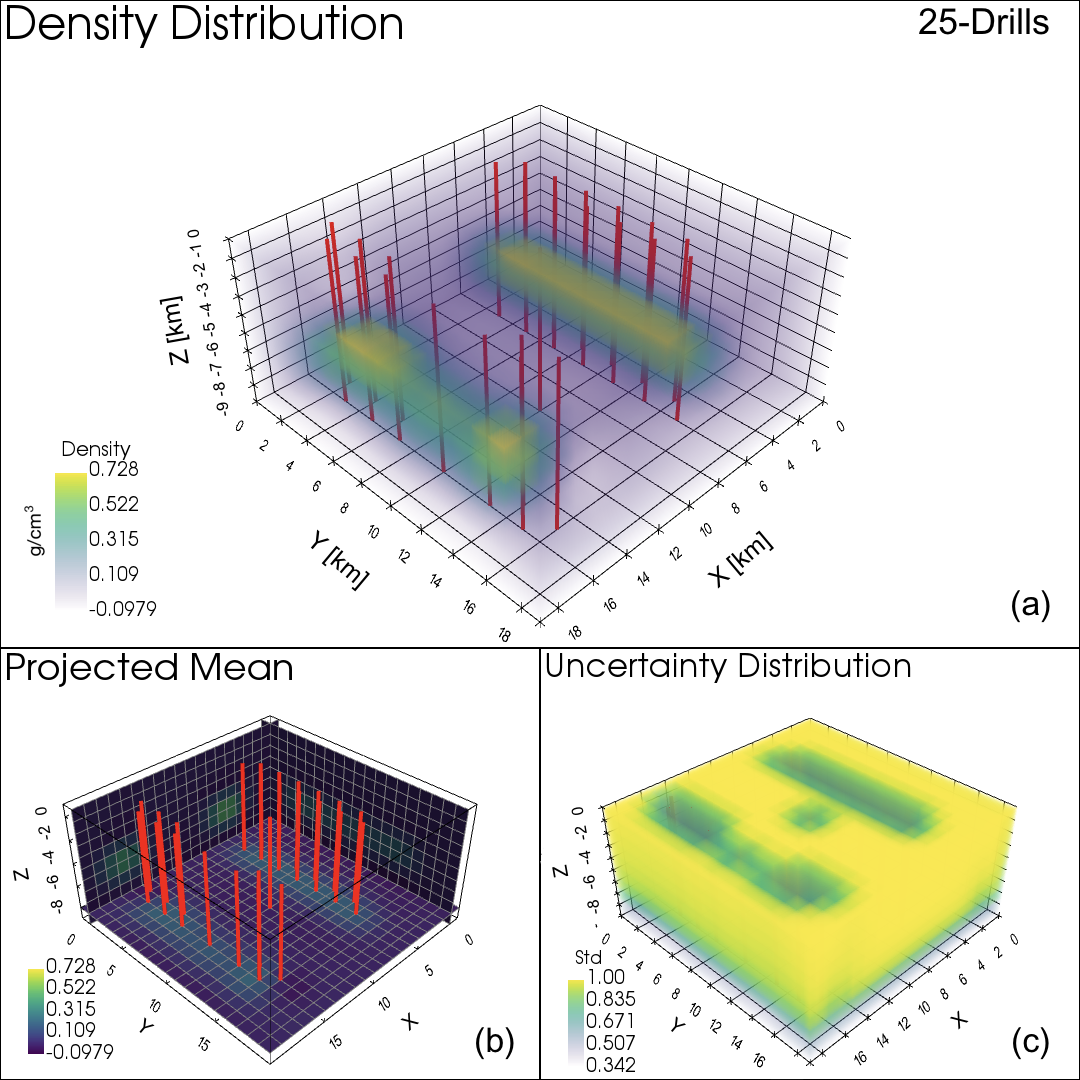}
\caption{The reconstructed density after 2, 8 , 16, and 25 BO steps for drill-core measurements based on the UCB method with $\kappa=1$. Sub-panels (a,b,c) are defined as in Figure 3.}
\label{fig_model1_4steps}
\end{figure}

We simulate exploration by drilling yielding a total of 25 drill-cores. The posterior of the geophysical properties, density and magnetic susceptibility, is updated and evaluated after each step of new drill-core data added. Figure~\ref{fig_model1_4steps} shows the reconstructed density model after 2, 8, 16, and 25 drill-cores. A video of the complete sequence is attached in the supplementary material.\par

\subsubsection{Dependence on Exploration-Exploitation Parameter $\kappa$}

\begin{figure}[!t]
\centering
\includegraphics[width=4.0in]{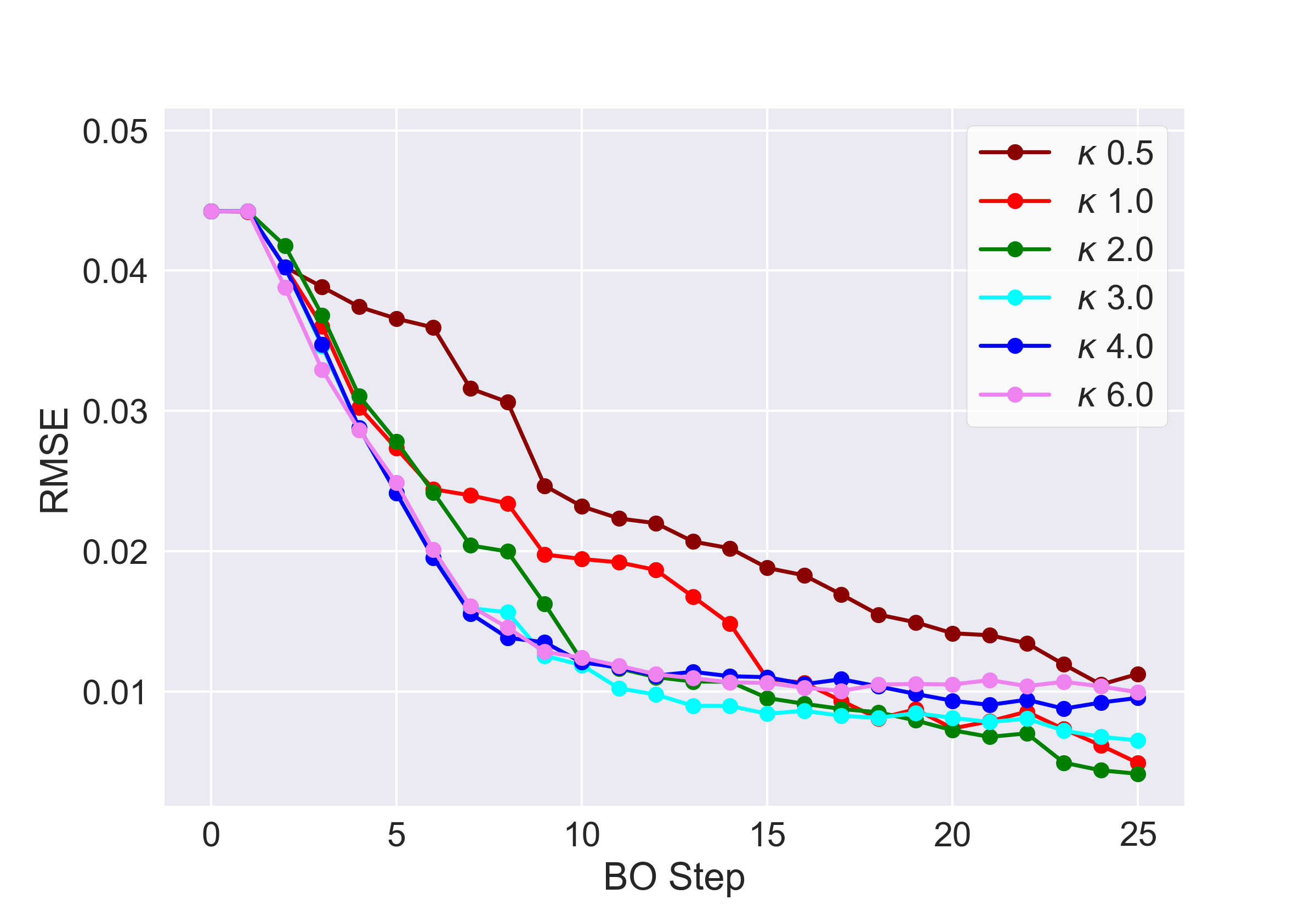}
\caption{The RMSE as function of drill-core number for a range of parameter $\kappa$.}
\label{fig_kappatest}
\end{figure}

We test the BO algorithm for a range of $\kappa$ values (see Equation \ref{eq_ucb}) in terms of their speed to converge towards the true model, as shown in Figure~\ref{fig_kappatest}. The difference between reconstructed and true model is given as Root Mean Square Error (RMSE) and becomes smaller with increasing number of drill-cores. The best results are typically achieved with a $\kappa$ of 1-3. While we kept $\kappa$ fixed for each BO experiment, in principle it is possible to employ an adaptive $\kappa$ that changes with the number of drill-cores (e.g., to, change from exploration to exploitation after a certain threshold of information gain is reached). Note that any parameter can be updated during the exploration. Moreover, it is possible to marginalize the posterior of the reconstructed model over a wide parameter range of GP length-scales, either using a uniform space or weighted by the corresponding marginalized log likelihood.\par

\subsubsection{Including Non-uniform Drill-Core Cost Function}

\begin{figure}[!t]
\centering
\includegraphics[width=2.6in]{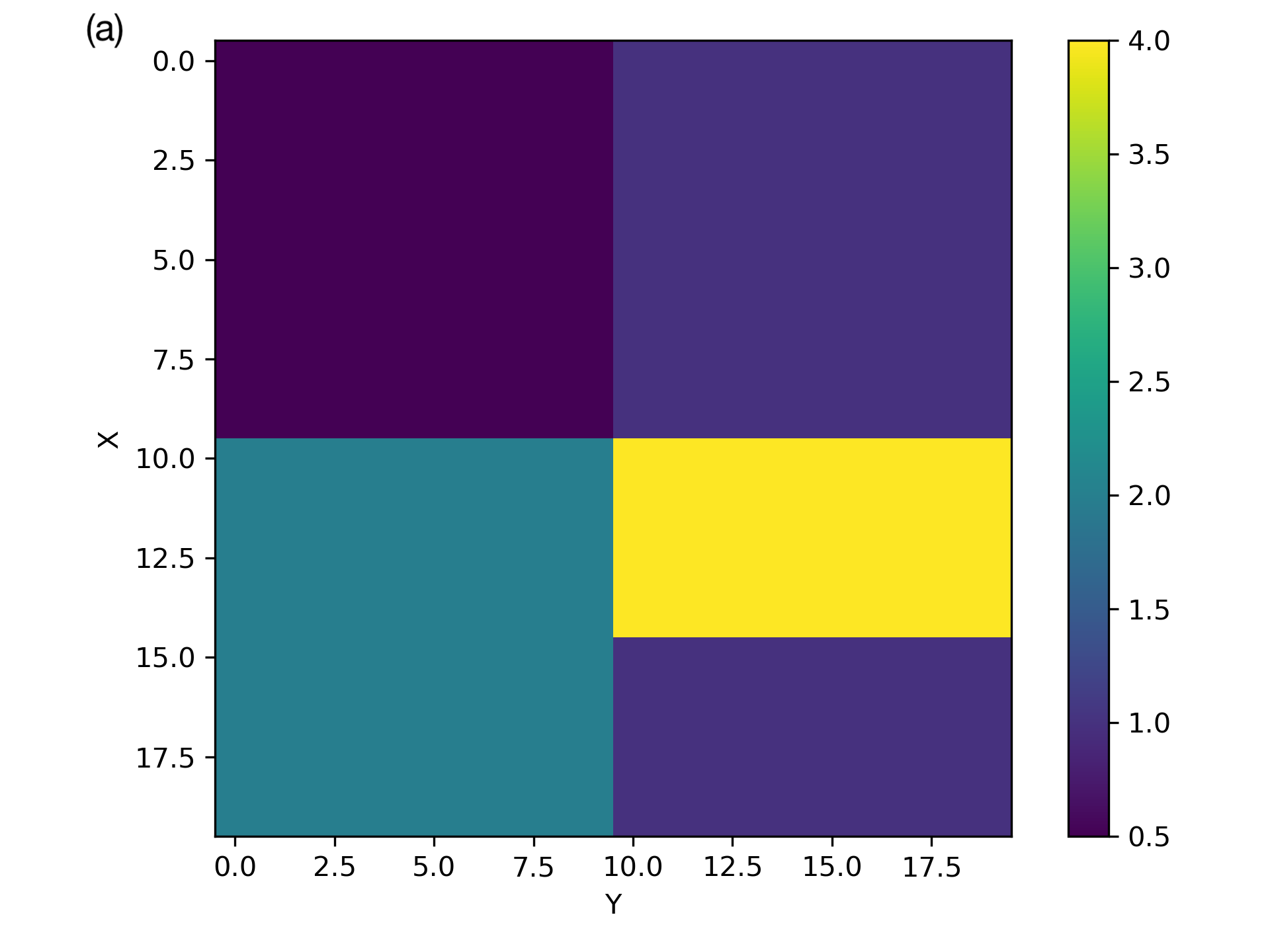}
\includegraphics[width=2.9in]{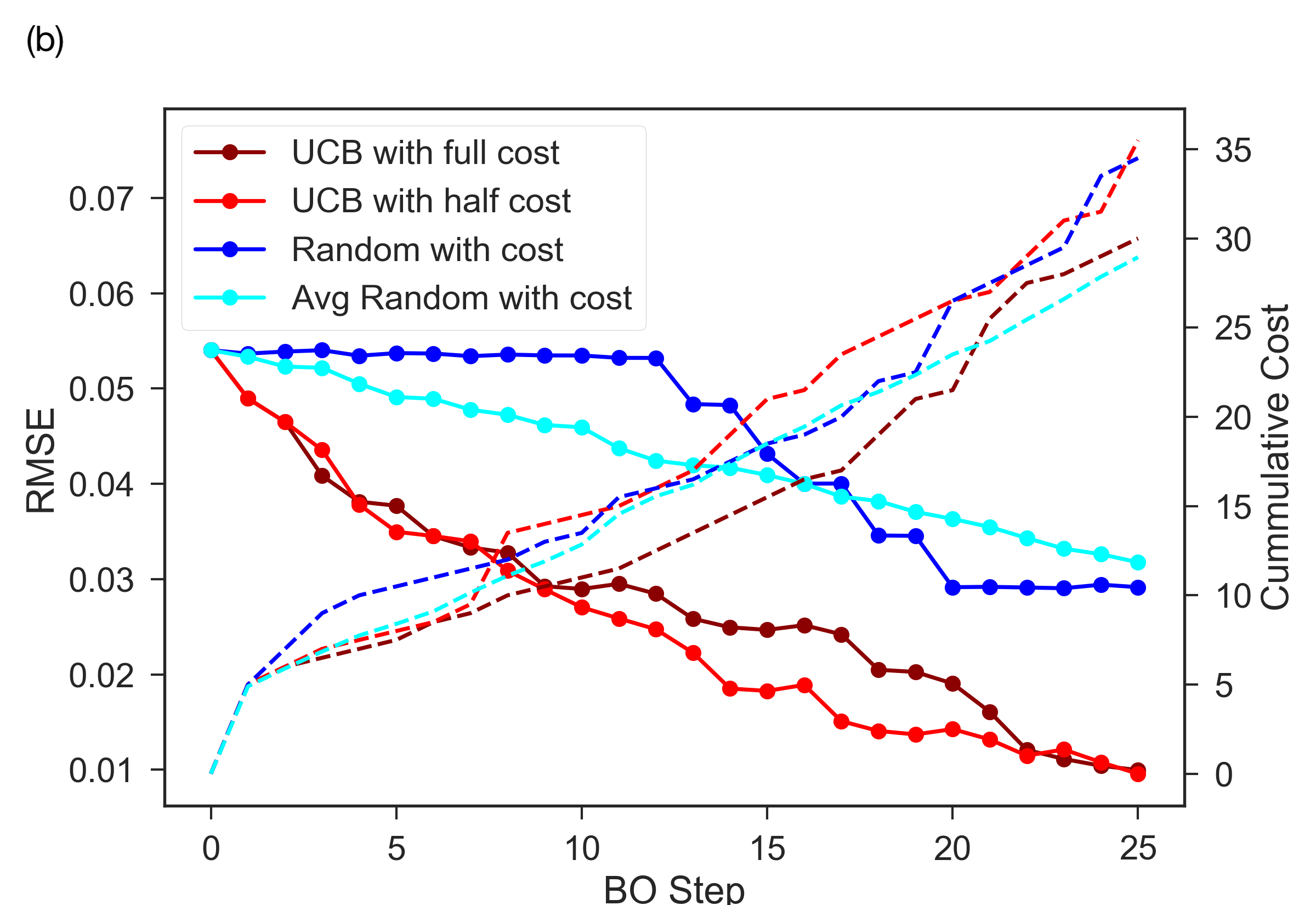}
\caption{(a) Example of a non-uniform cost function for drilling at different land locations, and (b) the results of three experiments with the same non-uniform cost function show how the UCB method (with two different cost weightings) compares to a standard non-BO method (see description in text) in terms of RMSE (solid line) and cumulative costs (dashed line).}
\label{fig_drillcost2D}
\end{figure}

In the following we test how our optimisation approach performs if we add a non-uniform cost-function that depends on the location price where the drill-core is sampled. Figure~\ref{fig_drillcost2D} shows an example of a spatial distribution of a cost-function, which can in practice, e.g., represent different drill-costs for farm land and unused land. We compare the results of the BO UCB method (equation~\ref{eq_ucb}) for two different weightings of costs ($\gamma = 0.5$ and $\gamma = 1.0$ ) to a standard non-BO method that relies on random drill-core sampling that is probabilistically weighted with the measured gravitation field divided by the cost at this location. In addition we show the average of ten such random drill-core samples. The experiment shows that the BO method with full cost weighting significantly outperforms the standard non-BO method at roughly the same cumulative costs of drilling (see Figure~\ref{fig_drillcost2D}b). The BO method with full cost weight shows only a slow increase in cost until a larger jump in cost is necessary to significantly reduce the RMSE further (BO step 17-18 and 20-22). If the objective function has a smaller weight on costs ( $\gamma = 0.5$ ), the RMSE reduces faster but at a higher price.

\subsection{Simulation Results: Multi-Layer Model}
\begin{figure}[!t]
\centering
\includegraphics[width=4.5in]{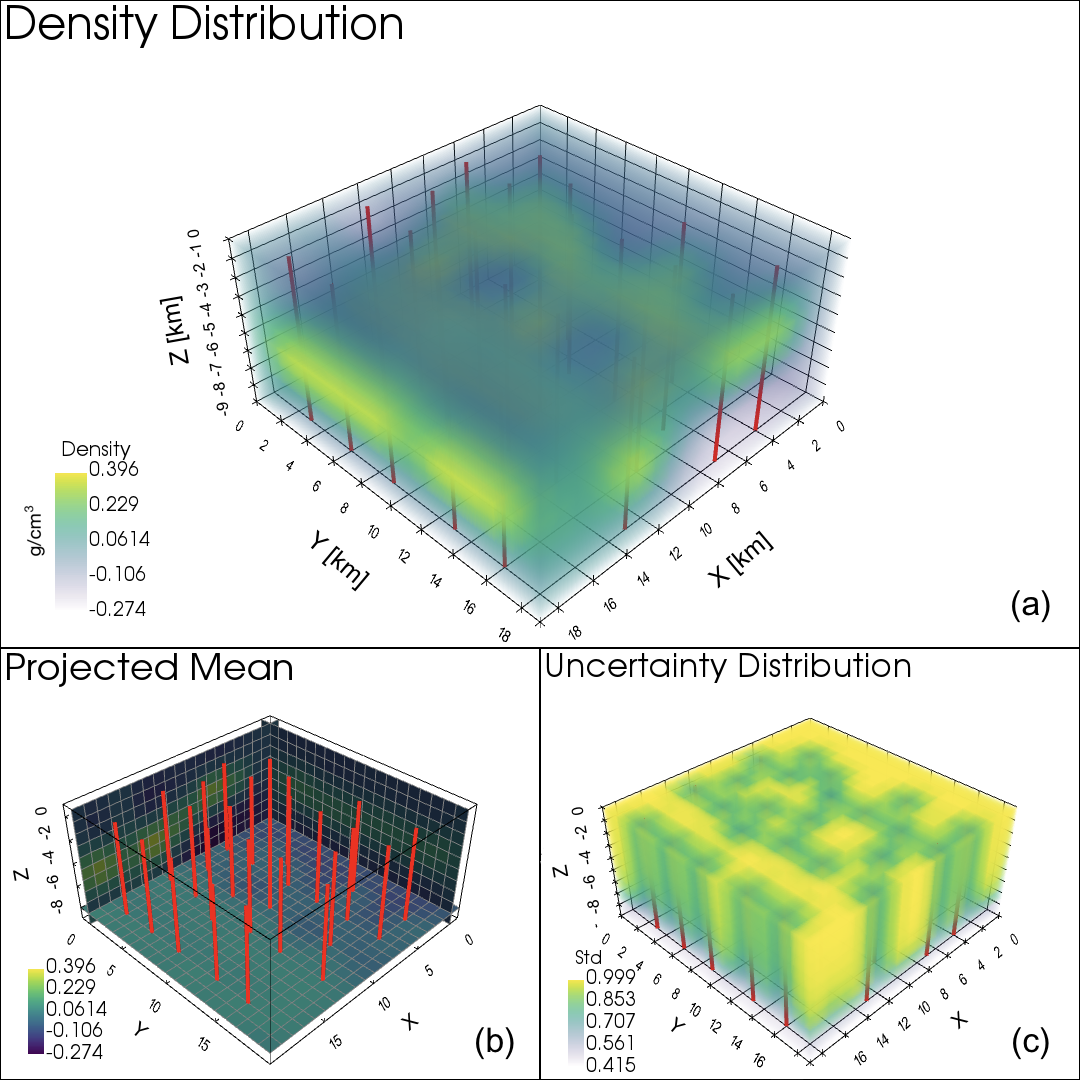}
\caption{The reconstructed density model of the layered model after 25 drill-core BO samples. Sub-panels (a,b,c) are defined as in Figure 3.}
\label{fig_model2_reconstructed}
\end{figure}

This synthetic model simulates a folded geological formation that consists of three layers with different density and magnetic susceptibility. The three layers are surrounded by a low density and magnetic medium. The layers are orientated parallel to the surface while being folded halfway in y-direction as shown in the middle panel of Figure~\ref{fig_modeloverview}.  Given that there is no clear separated concentration in density, the BO algorithm proposes drill-cores fairly uniformly, starting at the cube side with the lowest layer depth-to-surface and propagates drill-cores towards the other half of the cube. The reconstructed density distribution is shown in Figure\ref{fig_model2_reconstructed}.

\subsection{Simulation Results: Multi-Clump Model}
\begin{figure}[!t]
\centering
\includegraphics[width=4.5in]{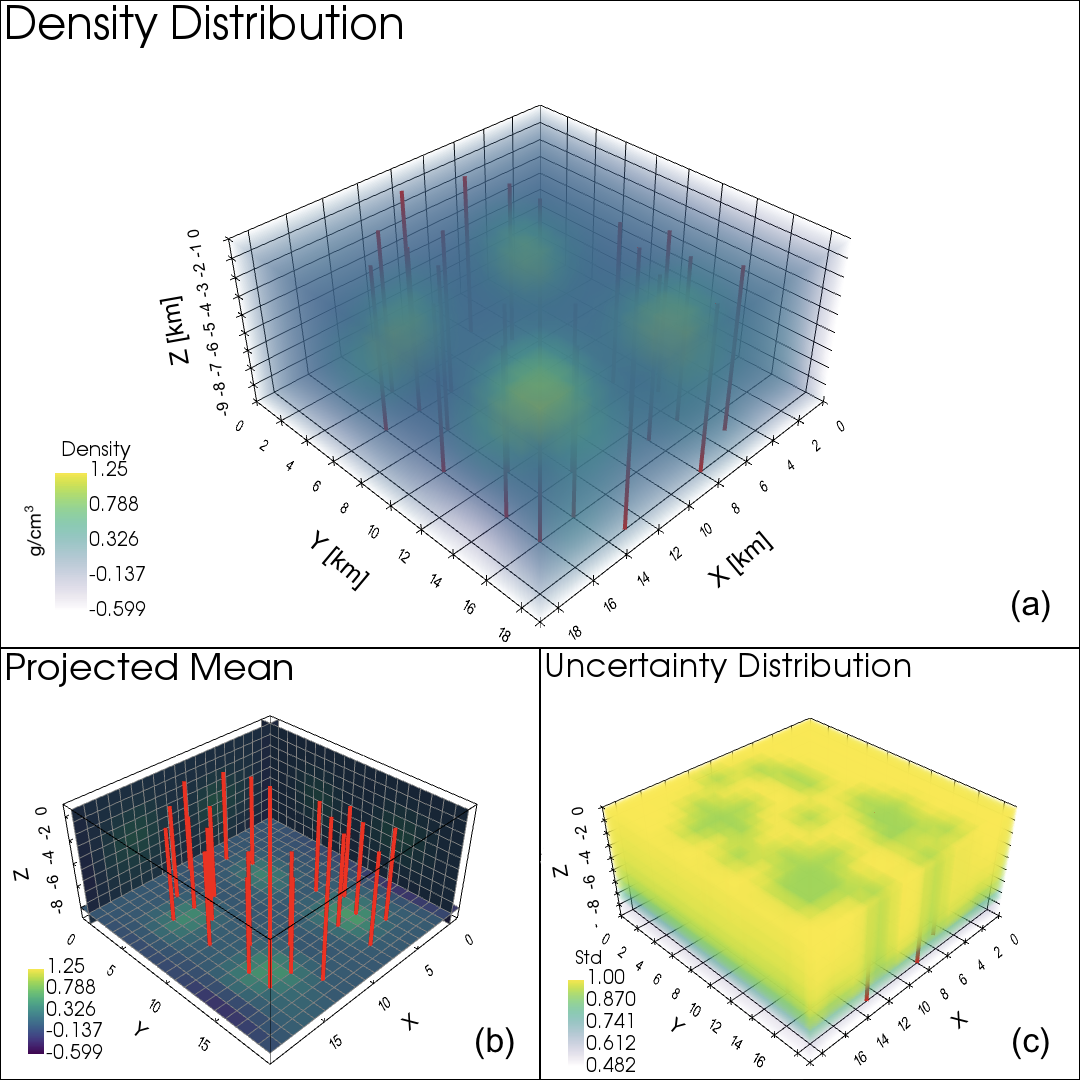}
\caption{The reconstructed density model of the clumpy model after 25 drill-core BO samples. Sub-panels (a,b,c) are defined as in Figure 3.}
\label{fig_model3_reconstructed}
\end{figure}

The clumpy model consists of four spherical high density and magnetic clumps that are well separated from each other as shown in bottom panel of Figure~\ref{fig_modeloverview}. Two of the four clumps are placed lower in vertical height than the other two. These two pairs are subdivided again in having different rock chemistry ($w_{D-M} = 1.0$ or $w_{D-M} = 0.5$).
The best reconstructed model, as based on the final RMSE after 25 drill-core sampling, is achieved using the UCB BO method. The reconstructed density distribution of this model is shown in Figure~\ref{fig_model3_reconstructed}.

\subsection{Evaluation of Optimisation Strategies}
\label{sec_eval}
\begin{figure}[!t]
\centering
\includegraphics[width=2.9in]{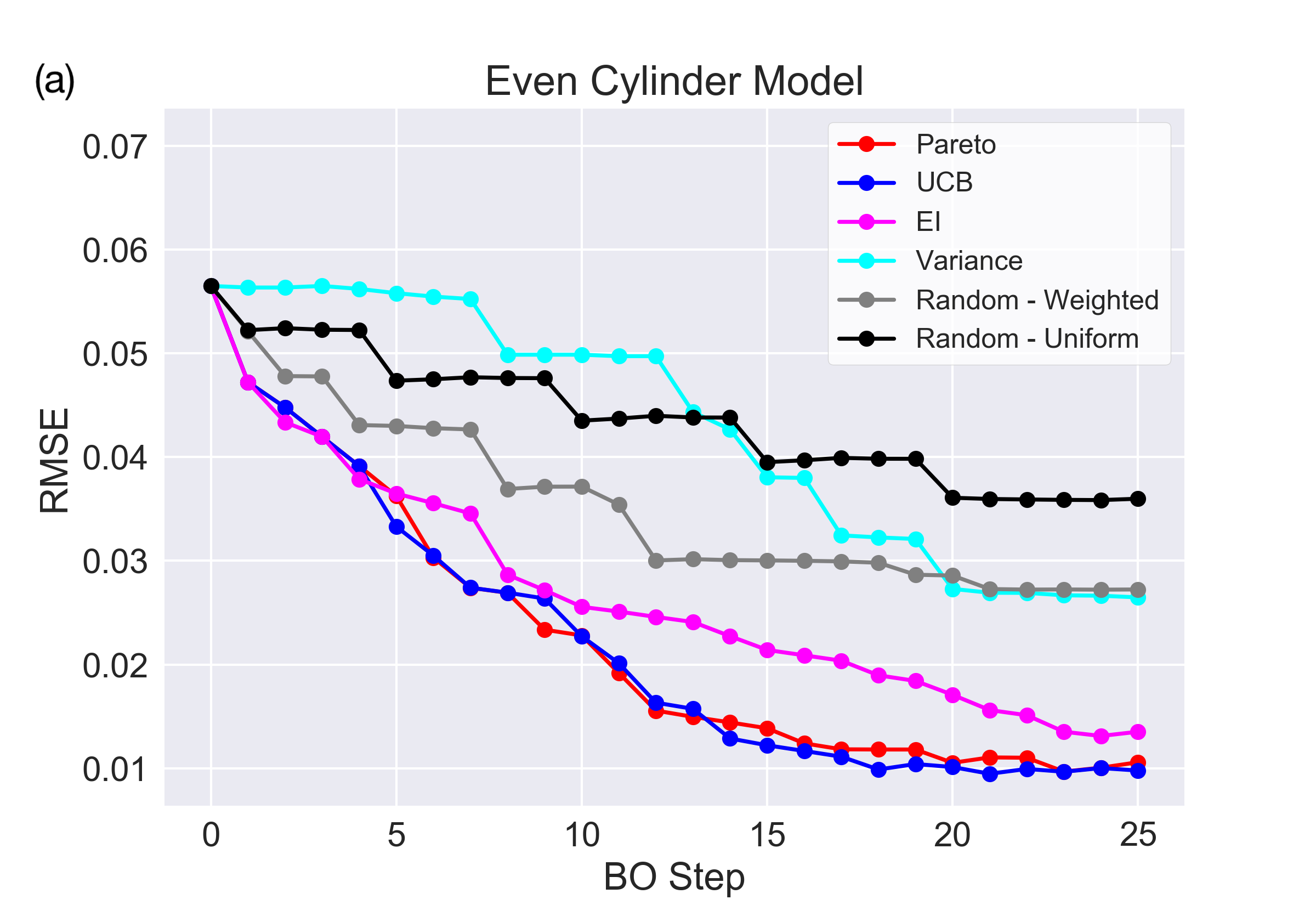}
\includegraphics[width=2.9in]{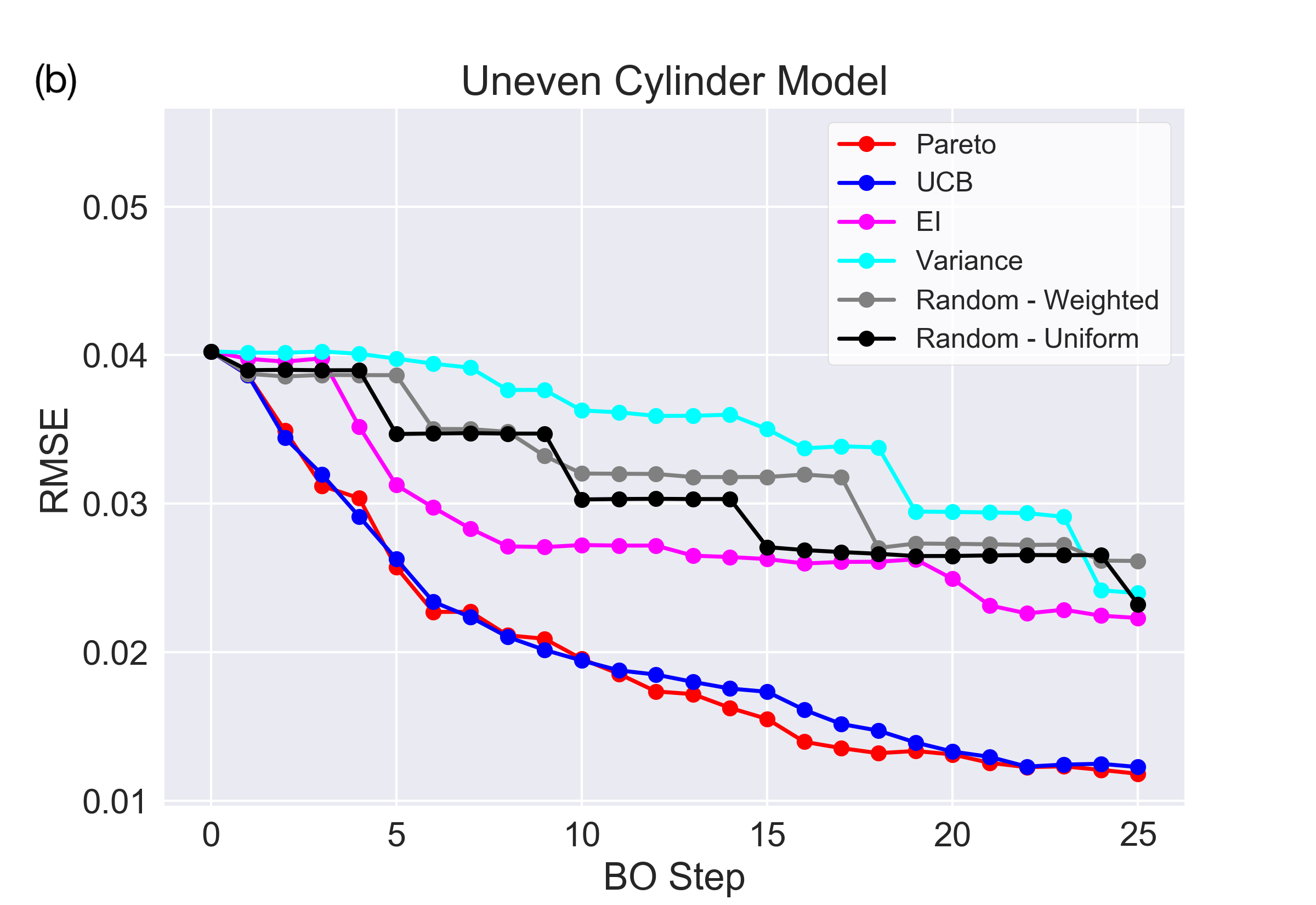}
\includegraphics[width=2.9in]{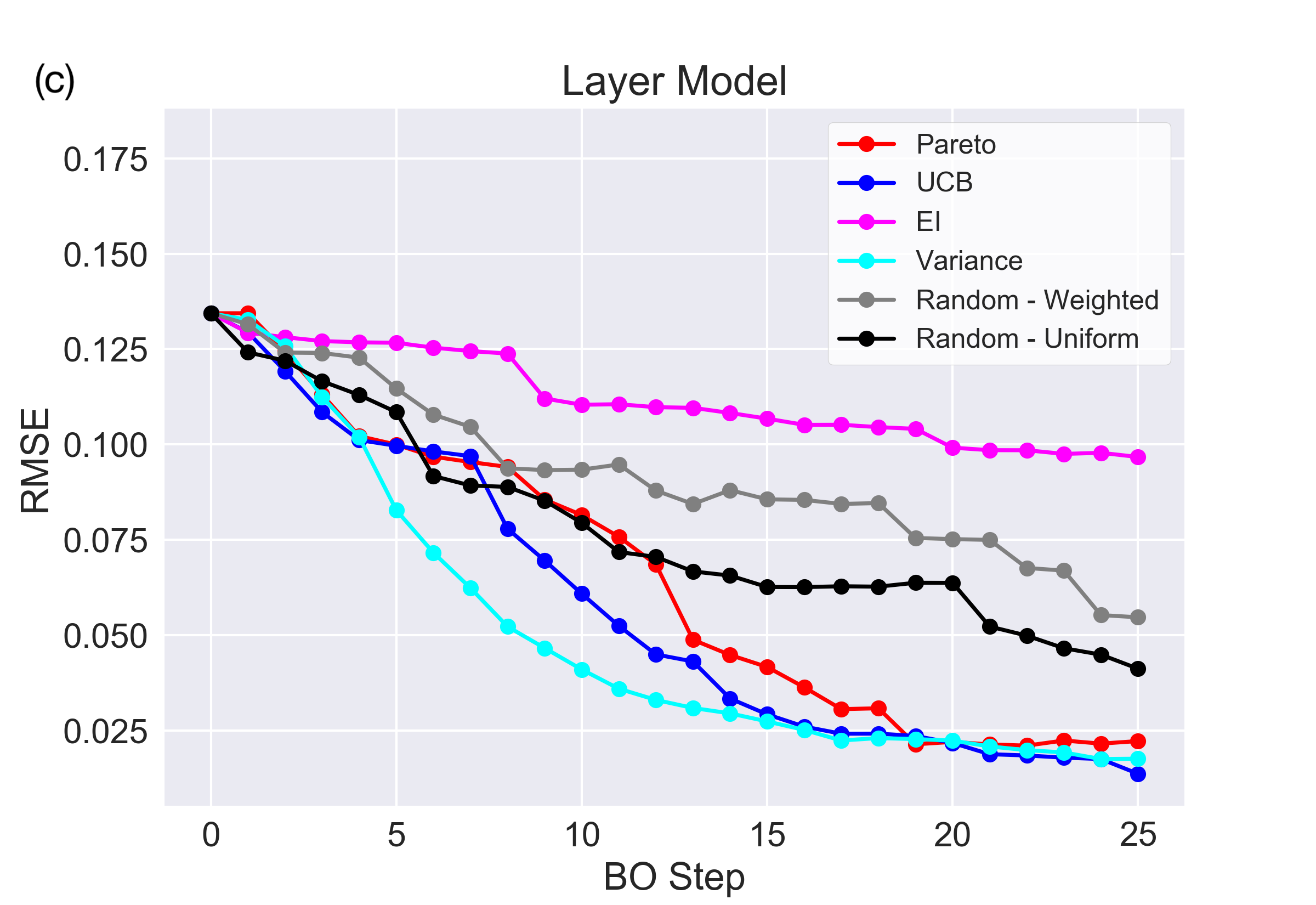}
\includegraphics[width=2.9in]{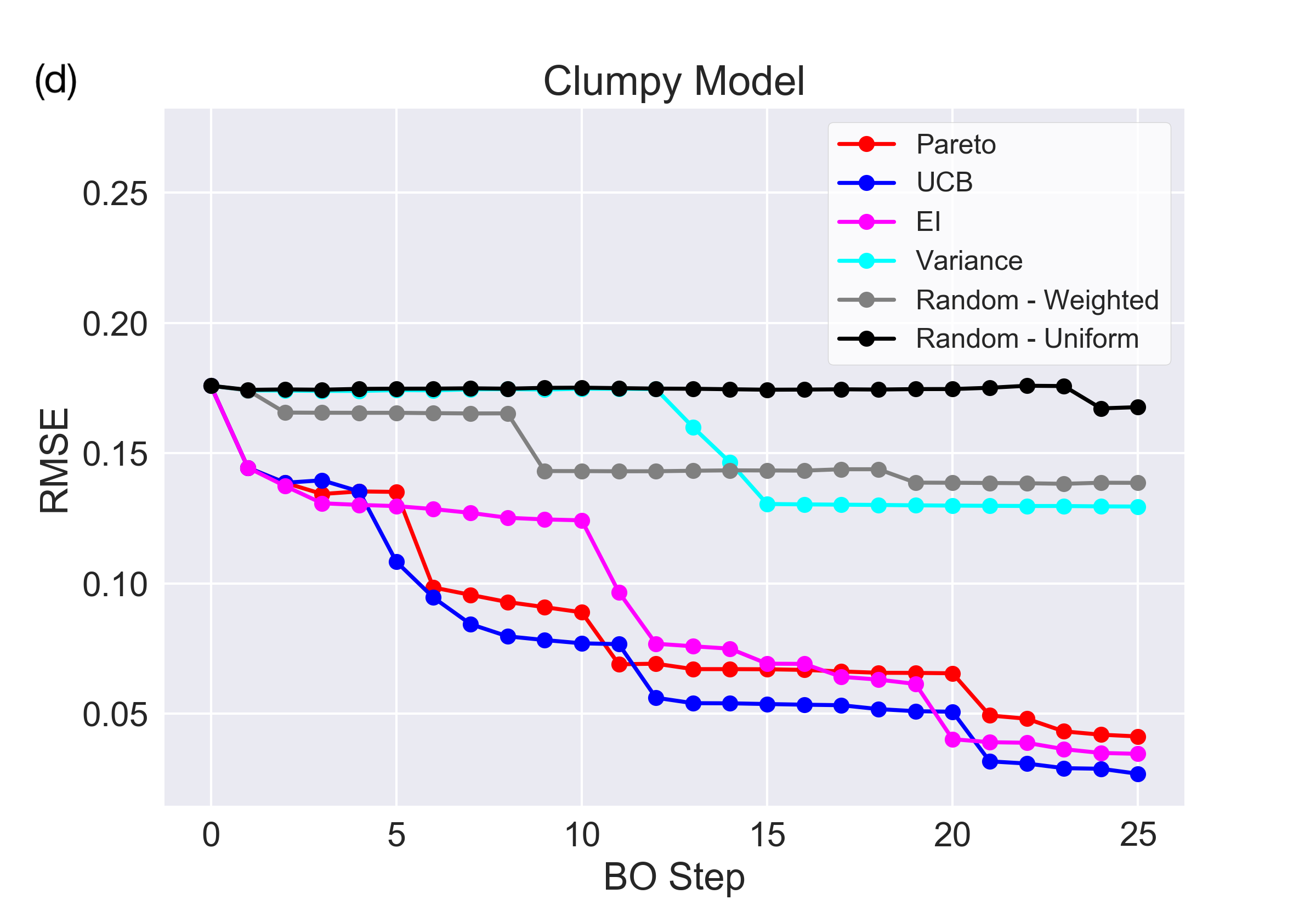}
\caption{Results of the RMSE values between reconstructed and true model as measured for different synthetic models: (a) even and (b) uneven cylinder model, (c) layer model, and (d) clumpy model, as described in text. Multiple BO drill-core selection strategies are compared against each other in terms of their performance for selecting additional drill-cores: 1) Pareto, 2) UCB, 3) Expected Improvement (EI), and 4) Variance only.  Additionally, two non-BO  standard methods are shown, which are 5) randomly sampling drill-cores from a uniform evenly spaced grid (Random - Uniform) and 6) random sampling with probabilistic sensor data weighting (Random - Weighted). In all four test-sets, Pareto and UCB converge faster towards the true model than the two standard methods.}
\label{fig_BOcomp}
\end{figure}

To evaluate the efficiency of the proposed BO  methods, we show in Figure~\ref{fig_BOcomp} the results of our simulated experiments for a sampling series of 25 drill-core locations. The RMSE values are calculated for the four different synthetic models (see model description above), and the performance of multiple BO functions are compared to two mock traditional non-BO approaches: The first one samples from a random selection of 25 uniform evenly spaced drill-core locations across the entire model space. The second non-BO method selects drill-core locations from a random distribution that is probabilistically weighted with the gravity and magnetic field sensor measurement. These two methods are compared to our BO using (1) Pareto EHVI, (2) the sum of the two UCB function over density and magnetic properties given as
\begin{equation}
\begin{split}
UCB (\mathbf { x }) = \mu_{density} ( \mathbf { x } ) + \kappa \cdot \sigma_{density} (\mathbf { x })  \\
+ \mu_{magnetic} +  ( \mathbf { x } ) + \kappa \cdot \sigma_{magnetic} (\mathbf { x }),
\end{split}
\end{equation}
(3) the expected improvement  \citep[EI,][]{mockus1978}, \cite{jones1998}, which is defined by maximizing the probability of improvement (PI) over the incumbent $f_{max}(x_{drill})$ with
\begin{equation}
   PI(x) = P(f(x) \geq (f_{max}(x_{drill})) = NormCDF \left[ \dfrac{\mu(x) - f_{max}(x_{drill})}{\sigma(x)} \right],
\end{equation}
given the normal cumulative distribution function $NormCDF$, and (4) optimising only the variance to minimize the total uncertainty of the reconstructed cube.

The results of these experiments are presented in Figure~\ref{fig_BOcomp}, which shows that the two BO strategies Pareto and UCB consistently outperform the non-BO methods by roughly a factor of two in final accuracy and speed of convergence towards the true model. Only in  case that the model consists of continuous smooth layers, the Variance-only BO method has the fastest convergence speed, but eventually converges for larger drill-core samples towards the same RMSE values as Pareto and UCB.

\section{Case Study on the Moomba Region}

\subsection{Data Measurements and Geometry}
\begin{figure}[!t]
\centering
\includegraphics[width=2.8in]{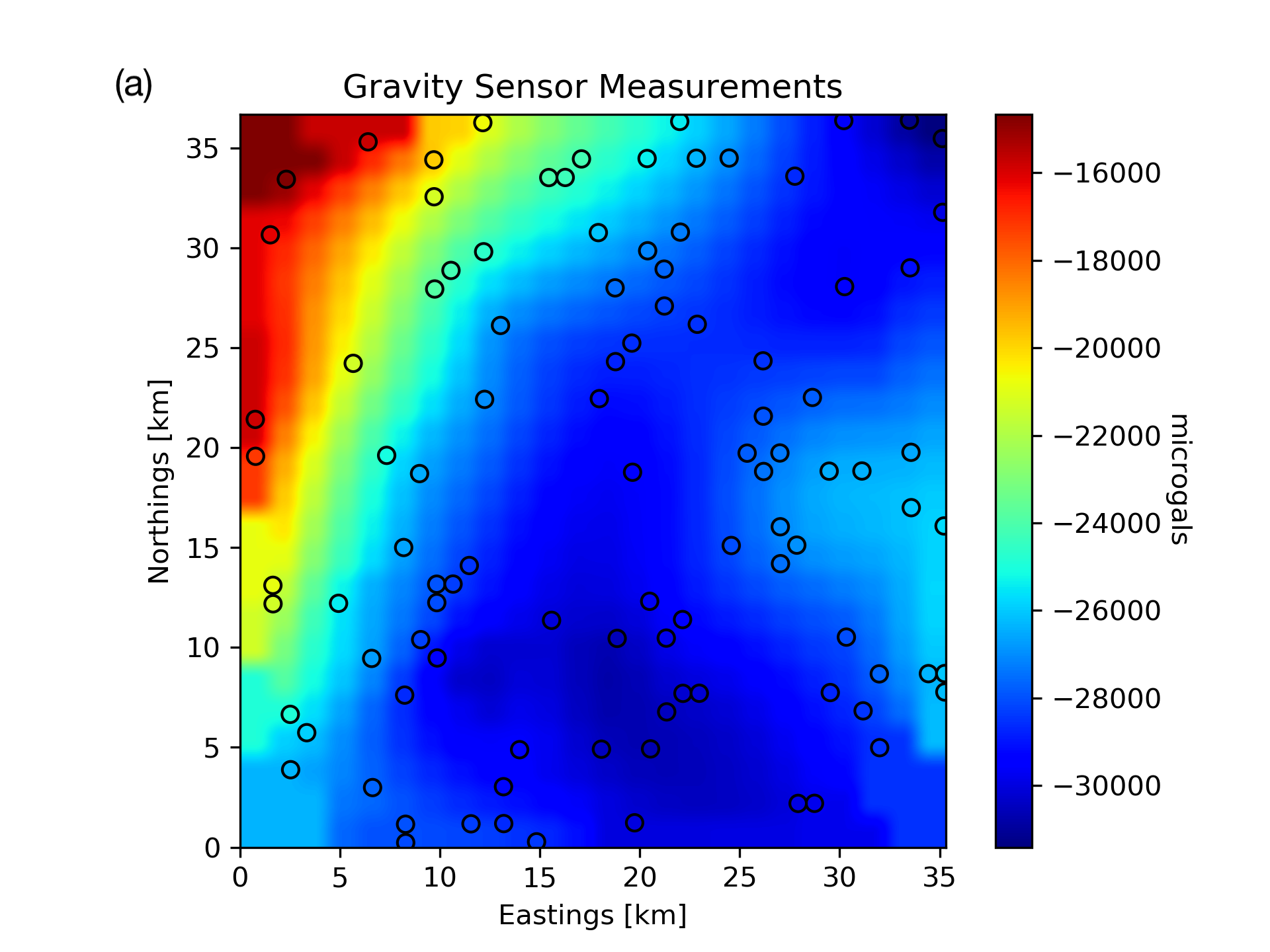}
\includegraphics[width=2.8in]{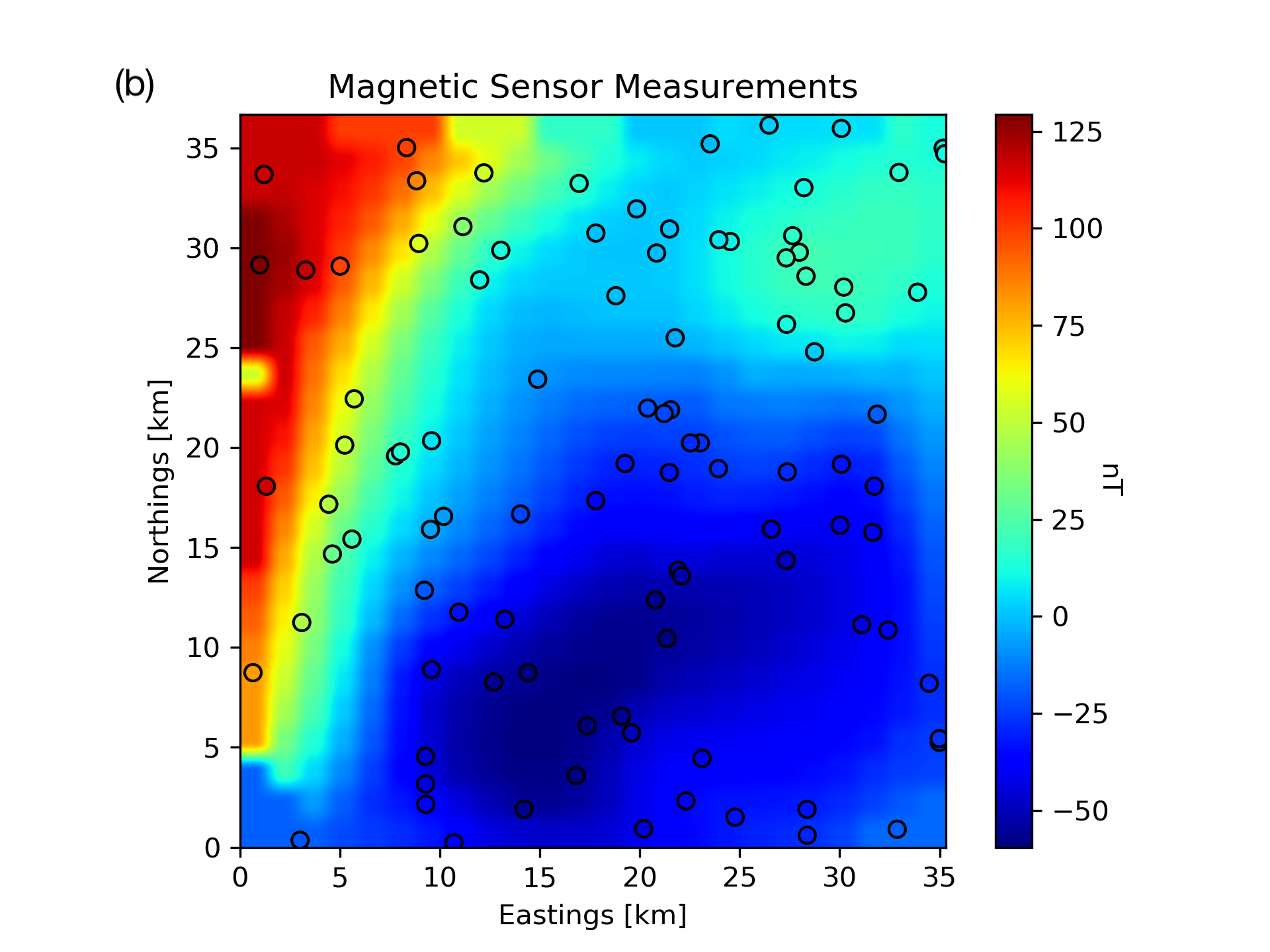}
\includegraphics[width=2.8in]{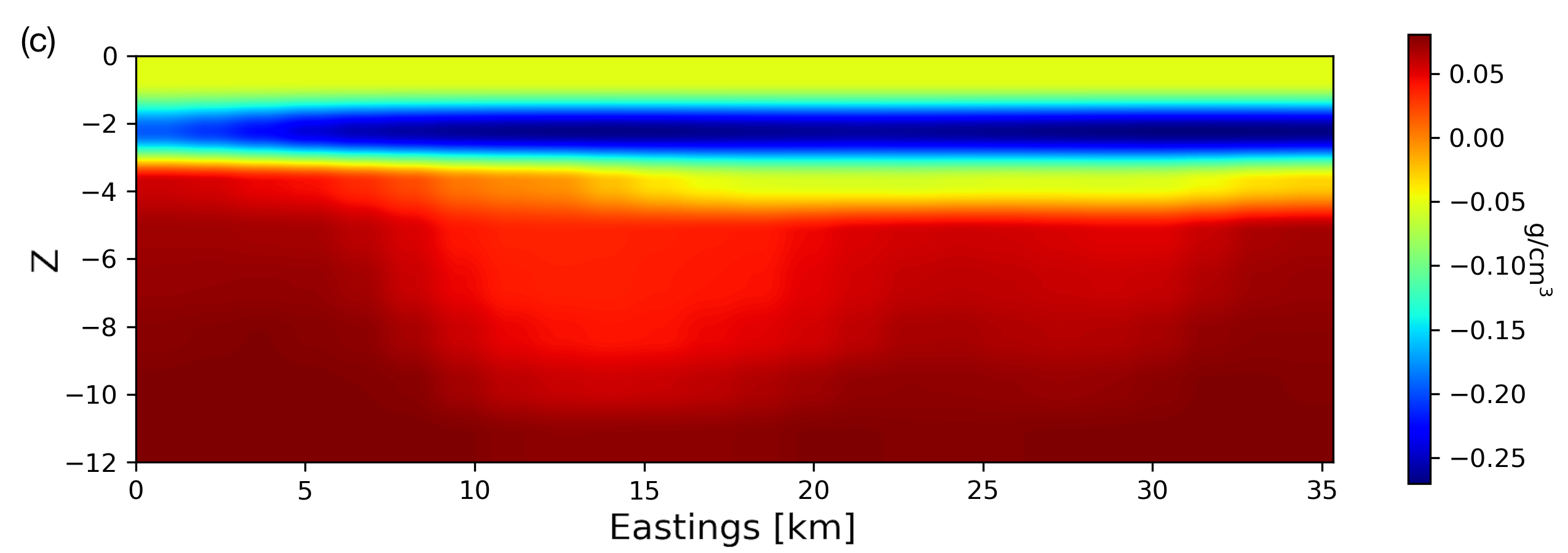}
\includegraphics[width=2.8in]{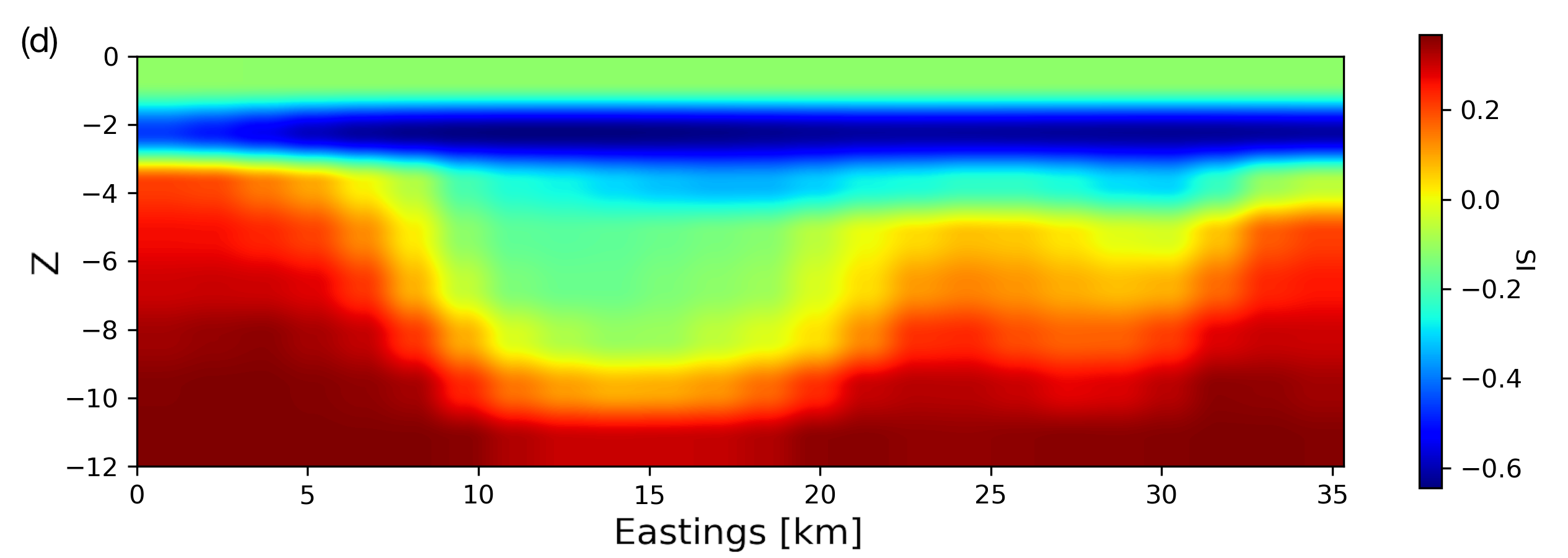}
\caption{(a) The interpolated gravity Bouguer and (b) magnetic anomaly field and sensor locations (circles) for the Moomba region. (c) the average vertical-profile of density and (d) magnetic susceptibility for the Moomba region as obtained from the mean of the Obsidian posterior distribution. Note that the central granite intrusion is less dense than the surrounding sediments and also has a lower magnetic susceptibility.}
\label{fig_moomba_sensors}
\end{figure}

To test the proposed joint BO method for a real geophysical formation scenario, we selected as test-set a basement in the vicinity of the Moomba gas field in NE South Australia. The survey data was collected over an area of 36 km $\times$ 36 km by 12 km deep, bounded by 410000-445000E, 6886000-6921000N. Regional sensor data are available in the form of Bouguer anomaly gravity and total magnetic field anomaly at a number of discrete locations and is interpolated to a uniform grid as shown in Figure~\ref{fig_moomba_sensors}. The model cube is divided into nx=24 by ny=24 by nz=8 cubes of dimension (1500m)$^3$ with a total number of 4608 voxels.\par

\subsection{Test Setup and Existing Geophysical Models}
To simulate and evaluate our bayesian optimisation strategy, we need to add drill-hole data after each BO step and evaluate the inversion results to the ground truth that holds the 3D geophysical properties. Since in practice it is not possible to obtain a complete large-scale 3D measurement cube of geophysical properties, we use instead the best available model as approximation of the true geophysical properties for the Moomba region in the southwestern portion of the Cooper Basin in South Australia. The portion of the Moomba region modeled here is characterized by a granite intrusion. Granites in this region are of economic interest for their geothermal energy potential and the associated sandstone-hosted Uranium deposits \citep[see][]{huston2011,khair2015}. A 3D model for the Moomba region has been generated previously with the Bayesian inference tool Obsidian \citep[see][]{obsidian2014,obsidian2016}, which uses Markov Chain Monte Carlo (MCMC) methods to sample from a joint probability distribution of rock properties for a set of parameterized layered models. Obsidian includes a range of forward models such as gravity, magnetic, 2D seismic reflection profiles as well as petroleum wells providing local constraints on formation depths and temperatures. It can incorporate prior knowledge of the geological structure and assumes a certain number of layers, whose rock properties and boundaries are constrained by Gaussian priors. Overall, it is currently the best 3D model available for this site.\par

For this test we take the mean of the final Obsidian posterior distribution over the 3D space as a proxy of the "true" basement model, i.e a granite intrusion surrounded by sedimentary deposits, from which we can sample drill-core measurements. Note that the true basement model is not seen by the BO selection algorithm and that only selected drill-hole data from this model are added to the inversion task along with the real gravity and magnetic anomaly field measurements for the Moomba region as show in Figure~\ref{fig_moomba_sensors}.   \par

\subsection{Experiment Results}

\begin{figure}[!t]
\centering
\includegraphics[width=2.7in]{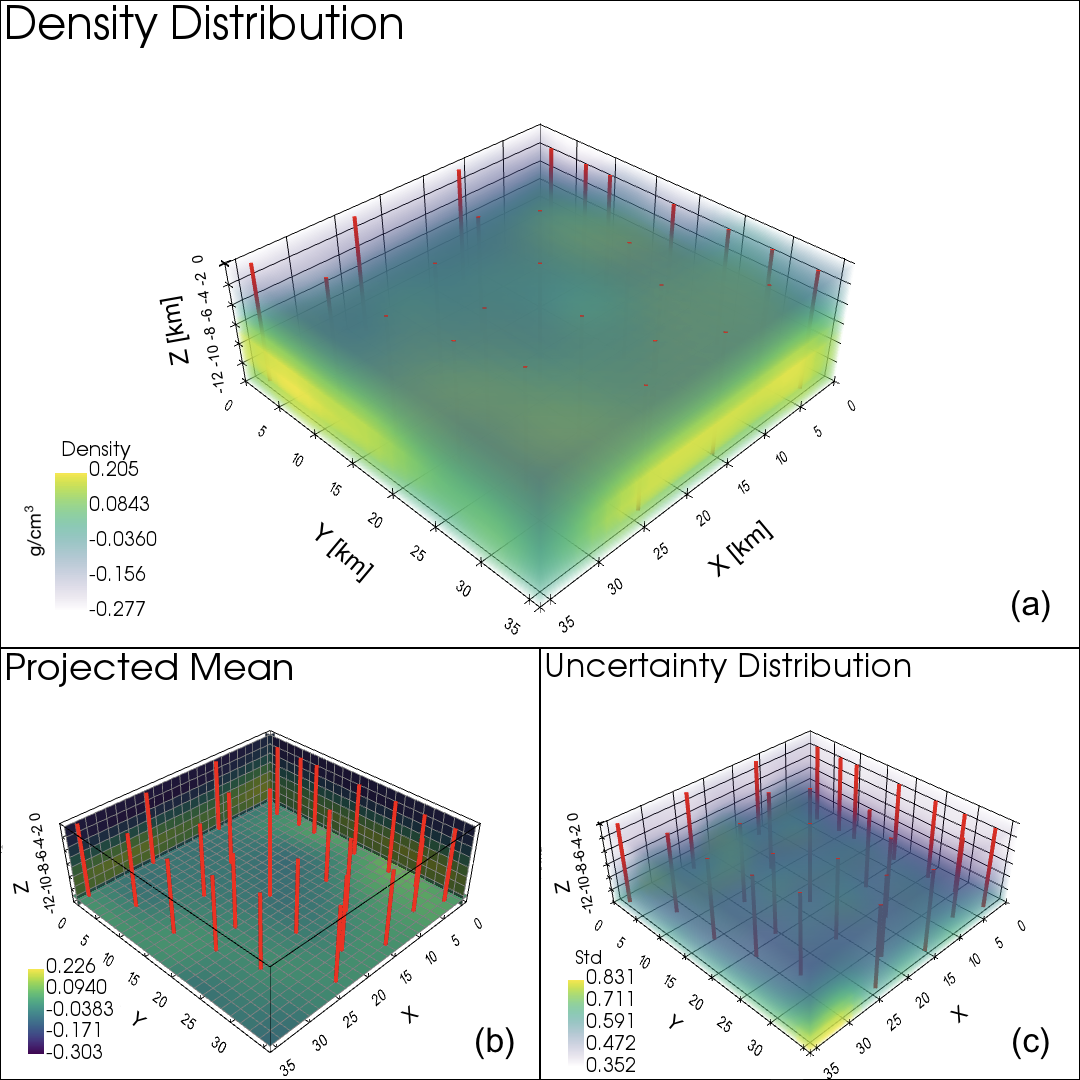}
\includegraphics[width=2.7in]{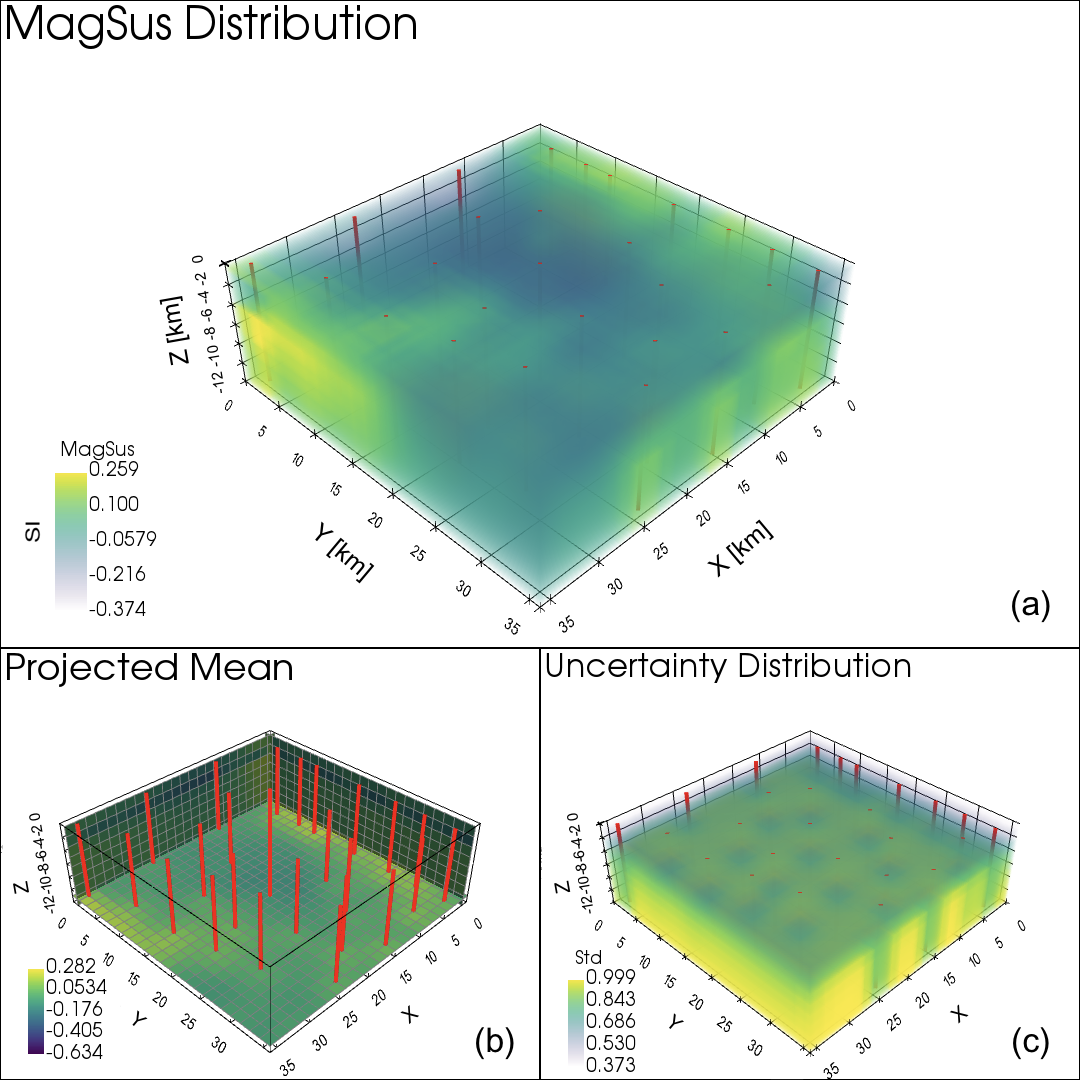}
\includegraphics[width=2.7in]{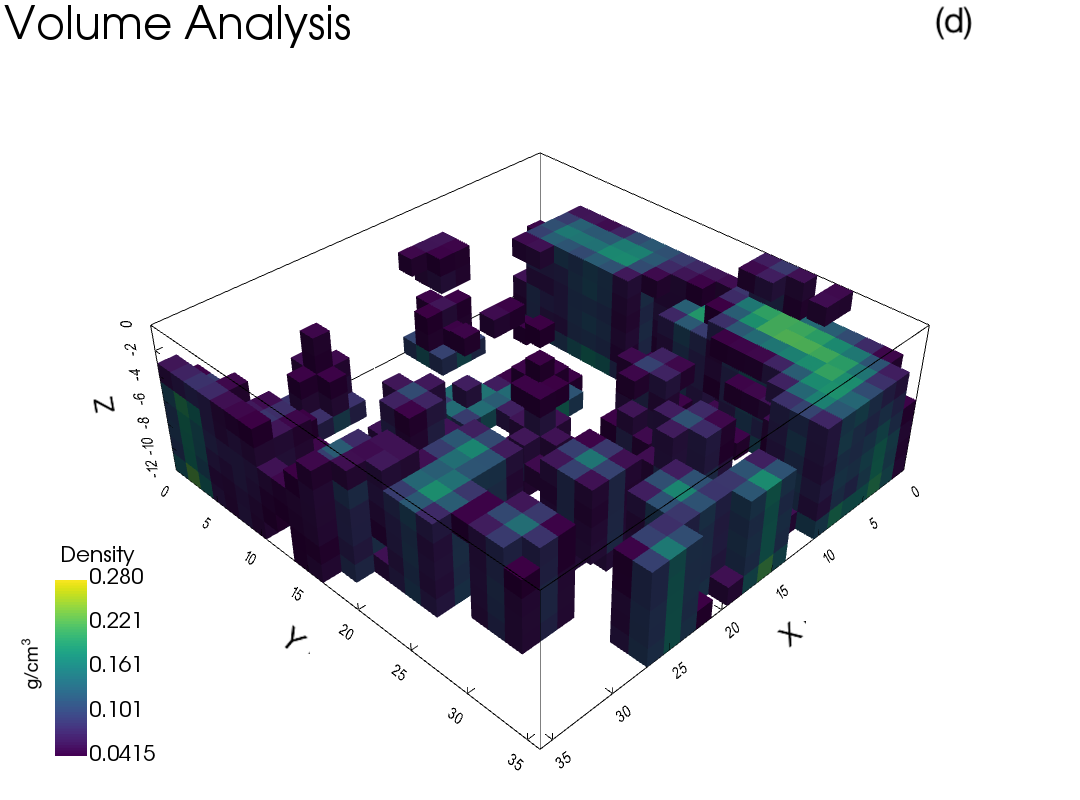}
\includegraphics[width=2.7in]{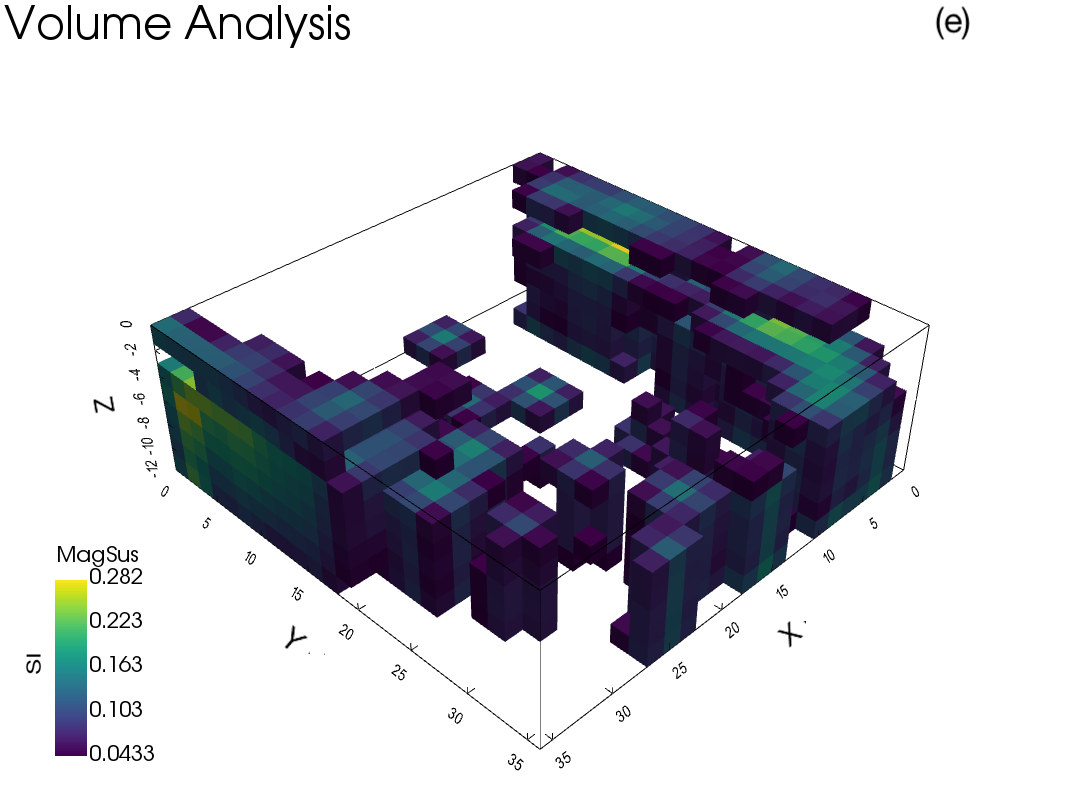}
\includegraphics[width=2.7in]{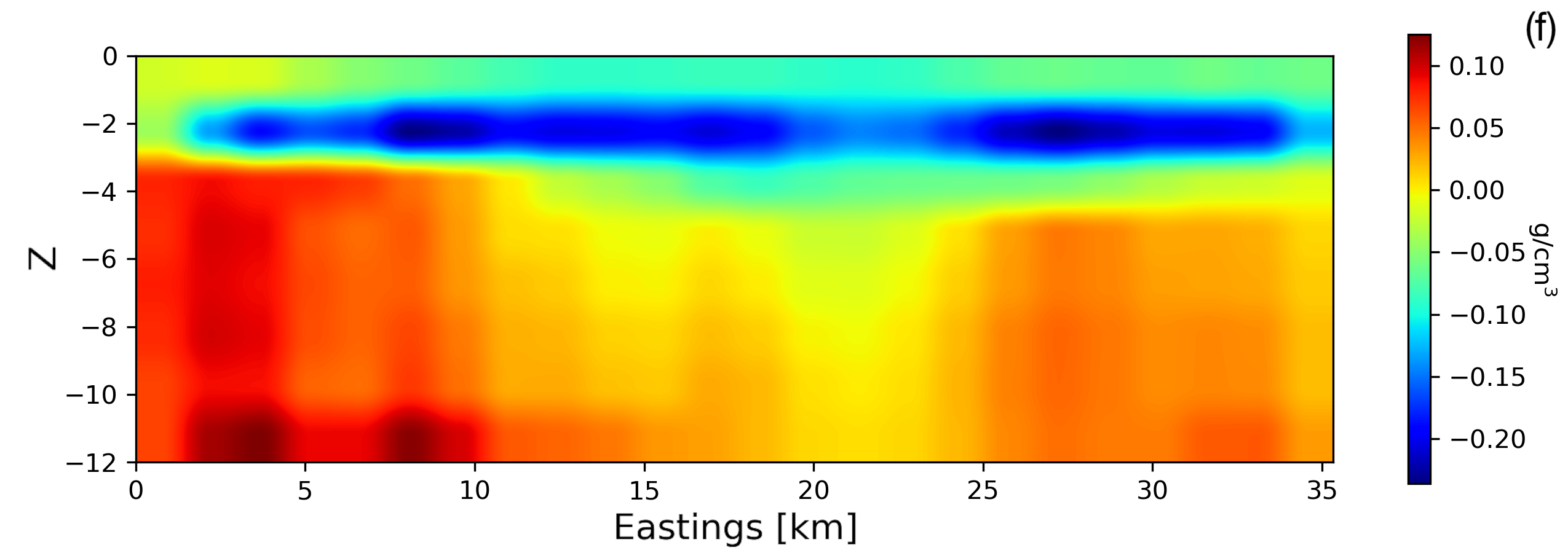}
\includegraphics[width=2.7in]{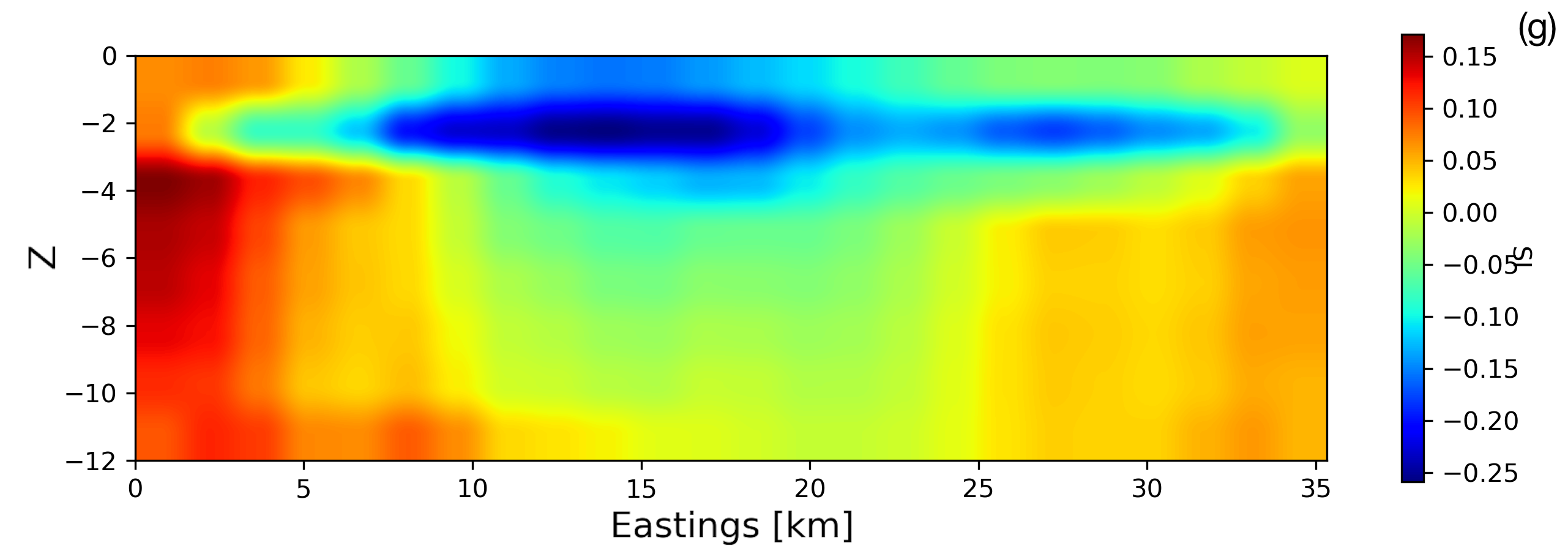}
\caption{The reconstructed density (Density Distribution) and magnetic susceptibility (MagSus Distribution) after 25 drill-core sampling. The distributions are shown after subtracting their mean value. Sub-panels (a,b,c) are defined as in Figure 3. The geometry is highlighted by the relatively high densities (d) and magnetic susceptibilities (e), forming a U-shape, reflect the sediments surrounding the central granite intrusion in this region. The average vertical-profiles are shown for the reconstructed density (f) and magnetic susceptibility (g).}
\label{fig_moomba_reconstructed}
\end{figure}

The simulated BO exploration experiment queries locations for new drill-cores given the gravity magnetic data and drill-core measurements obtained from the 3D Obsidian model. In total, 25 vertical drill-cores are sampled iteratively before the experiment stops. The sampling scheme and process of reconstructing the density and magnetic properties as function of drill-core is shown as videos in the supplementary material. The final reconstructed density and magnetic properties are shown in Figure~\ref{fig_moomba_reconstructed} after 25 drill-core measurements.\par

\begin{figure}[!t]
\centering
\includegraphics[width=2.9in]{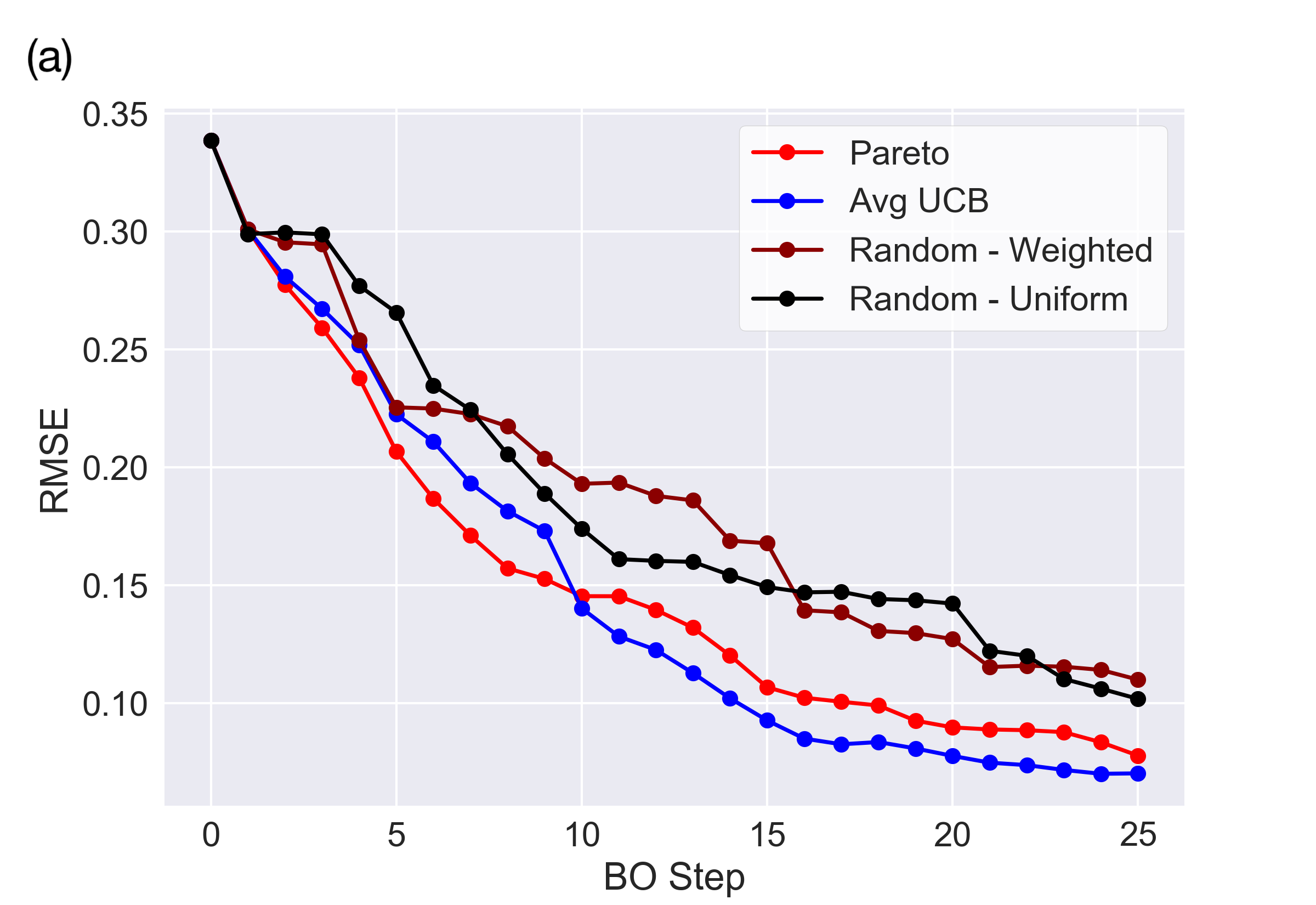}
\includegraphics[width=2.9in]{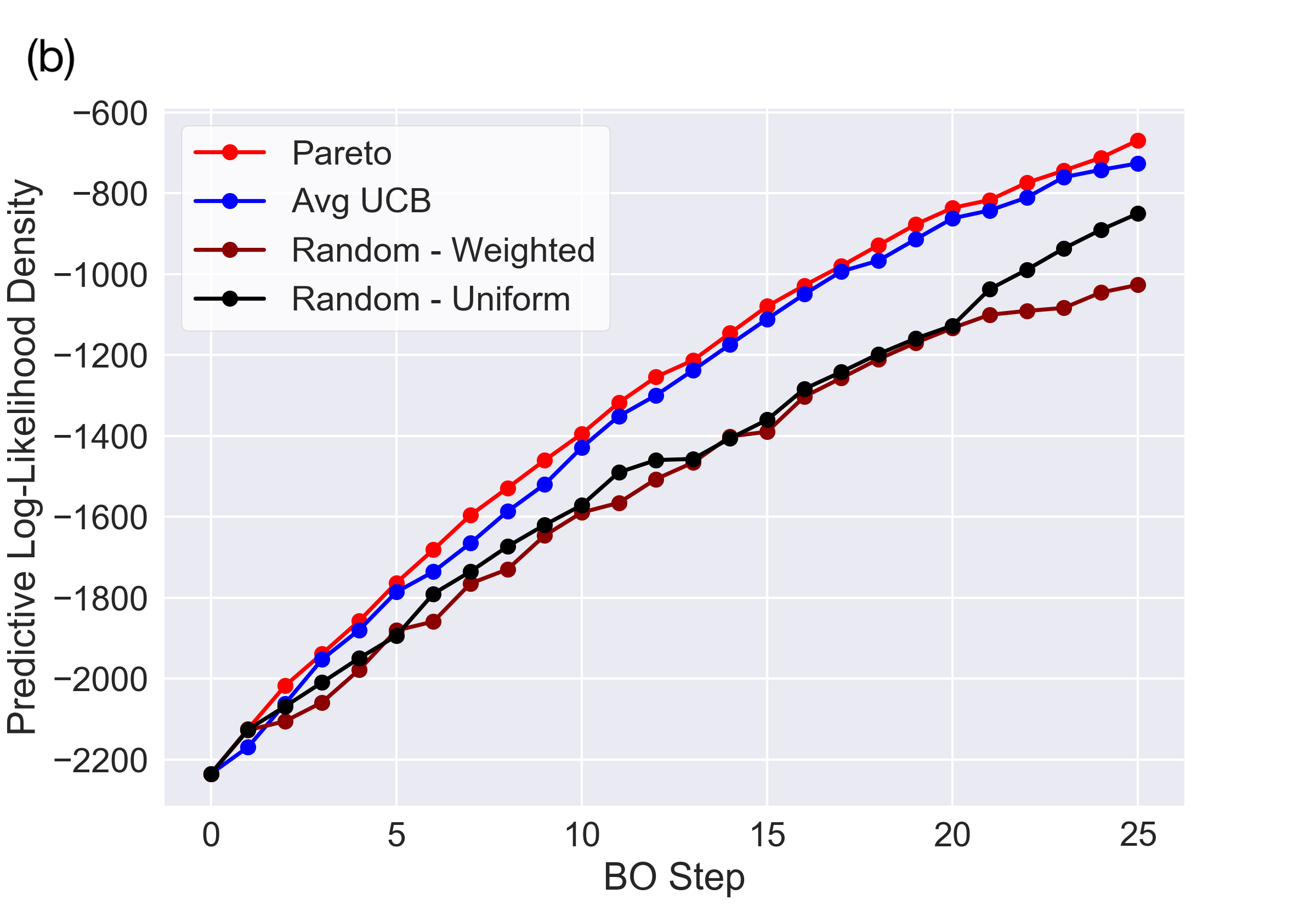}
\includegraphics[width=2.9in]{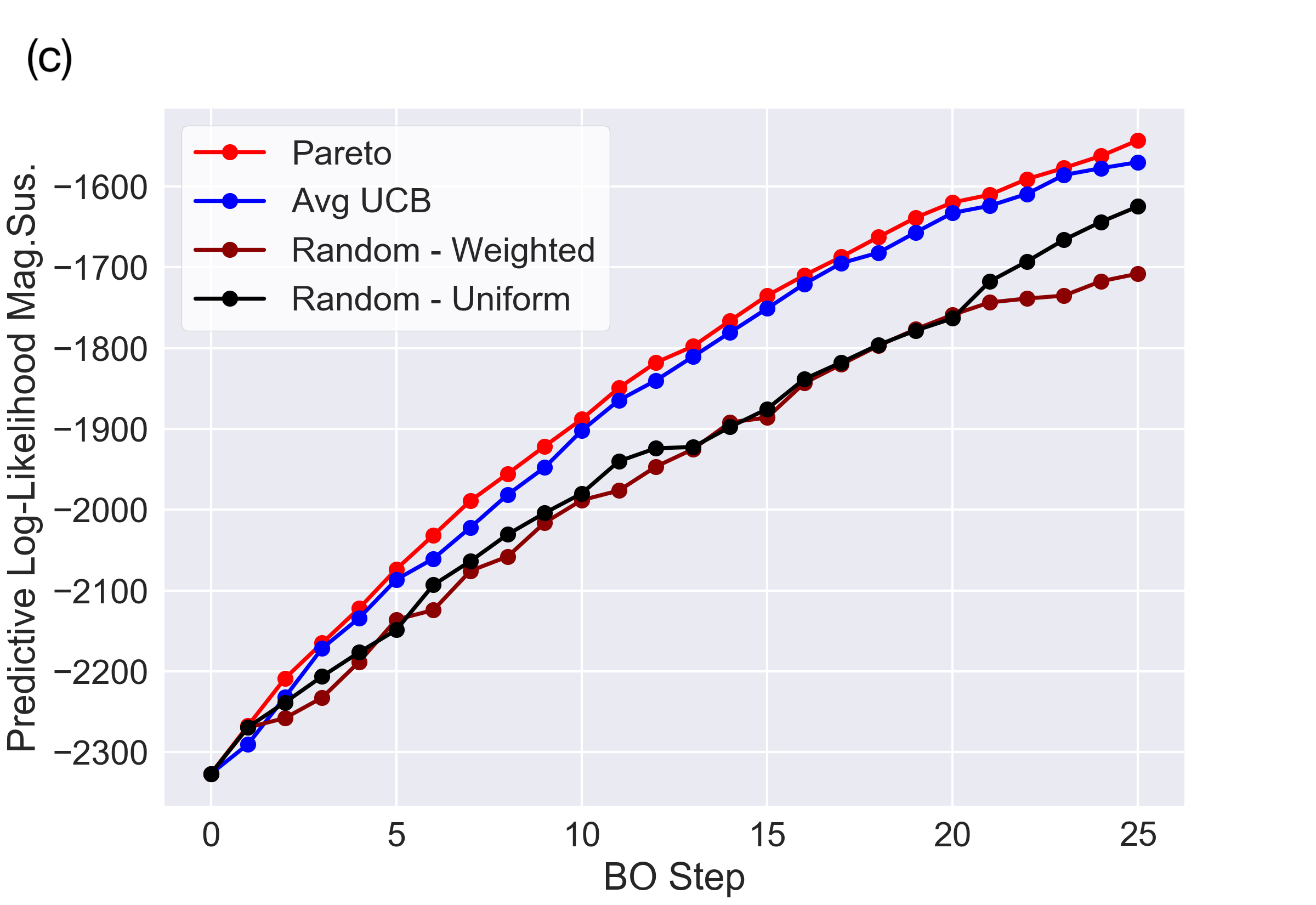}
\caption{Comparison between BO methods and random sampling as function of sampled drill-core number for the Moomba region: (a) total RMSE, (b) predictive log-likelihood (see equation \ref{eq_predloglike}, which describes the sum over all voxel probabilities considering the difference between observed and predicted mean values and uncertainties) for density, and (c) the predictive log-likelihood for magnetic susceptibility.}
\label{fig_moomba_BOeval}
\end{figure}

The experiment evaluates multiple optimisation schemes and calculates (1) the total RMSE between reconstructed and the true basement model as gauge for accuracy, and (2) the predictive log-likelihood, which considers also the predicted uncertainty $\sigma_i$ (square root of diagonal elements of the predicted covariance matrix, see equation 14),
\begin{eqnarray}
\ln L &=& \prod_{i=1}^N \ln \Big[ \dfrac{1}{\sqrt{2\pi \sigma_i^2}} \exp \big( \dfrac{-(y_i - y_i^*)^2}{2 \sigma_i^2} \big) \Big] \\
 &=& -0.5 \sum_{i=1}^N  \Big[ \ln (2 \pi \sigma_i^2)  + \dfrac{(y_i - y_i^*)^2}{\sigma_i^2}  \Big],
\label{eq_predloglike}
\end{eqnarray}
where $(y_i - y_i^*)$ is the residual between ground truth and the reconstructed mean predictive value for each voxel $i$. The performance evaluation is shown in Figure~\ref{fig_moomba_BOeval} for the two BO methods, Pareto and UCB, and compared to the two random based sampling methods. The results show that both BO methods outperform the non-BO methods in terms of accuracy, effectiveness, and maximizing the predictive log-likelihood.

\section{Discussion}
\label{sec_discussion}

In this paper we develop a probabilistic framework for joint inversion and multi-objective BO to select new sensor measurements, and demonstrate its capabilities on a test-set of synthetic geological models as well as real world scenario by solving the example problem of allocating iteratively new expensive drill-core samples. Our results show that this method can accurately reconstruct from sparse 2D sensor data a 3D model of multiple geophysical properties  and their associated uncertainties. Multiple BO sampling schemes are evaluated in terms of accuracy and effectiveness for a wide range of geological models and compared to traditional non-BO sampling schemes. The results demonstrate that BO can significantly outperform traditional non-BO sampling schemes, which may ultimately lead to more efficient and cost-saving sensor selection.\par

A key feature of the applied method is to consider all cross-variances between geophysical properties by solving simultaneously multiple forward models and to fuse jointly multiple and distinct sensors. Thus, this approach can take full advantage of additional sensor information to reveal properties of the reconstructed geophysical model at higher accuracy. Moreover, by using non-parametric Gaussian Process priors, the solution of the inverse problem provides a complete posterior distribution for all geophysical properties under consideration at any location or grid resolution. The reconstructed geophysical properties are described by a mean and variance value at each location (voxel), which are the two key-ingredients for Bayesian optimisation.\par

The proposed model allows applications to specify priors as additional options: (1) the typical length-scale of the geological structure or the spatial resolution of the reconstructed model via GP hyperparameters, (2) the gain/cost parameters in the BO acquisition function, and (3) the correlation coefficients between different geophysical properties. If the application is purely data driven (default option), GP hyperparameters and correlation coefficients are optimized given equation \ref{eq_logl}. Within the BO sampling iteration process, all parameters are adaptive towards possible changes in objectives over time. Alternatively, in case the aim is to reduce the dependency on certain prior parameters, it is straightforward to marginalize over the corresponding BO or GP hyperparameters, which increases the computational time but can offer a more accurate posterior distribution.\par 

Further improvement in accuracy and computational performance  can be achieved by dividing the cube in multiple cube elements that are larger than the maximal spatial correlation length-scale. Solving the inversion step for individual cube elements has the following advantages: (1) to optimise the GP hyperparameter length-scale specifically for each stationary dependent cube element (e.g., for tracing different geological structures), and (2) to update only the cube element(s) where new data are added. Note that the BO will still select the optimal new sensor data from the combined cube but the computational time for optimizing the acquisition function is much smaller in comparison to computing the GP inversion.\par

While the proposed method of inversion via GP is limited to linear forward models, such as gravity and magnetics, the reconstructed 3D distribution of the geophysical properties can be easily refined further by other probabilistic methods. This subsequent modeling can, e.g., take into account site-specific geological knowledge as well as other measurements that require non-linear forward models (e.g., seismic data), which rely on sampling from model priors or a distribution of possible models rather than solving the inversion directly. Such sampling models \citep[e.g., Obsidian or GemPy, see][ respectively]{shamsipour2012, obsidian2016} are computationally very expensive, even after considering MCMC sampling or variational inference, given their need to sample over an almost infinite range of possible models, and are therefore in practice typically constrained by certain assumptions about the underlying geological formation and  prior parameterisation. In these cases, the proposed GP method in this paper may significantly help to boost the sampling efficiency by providing the probability distribution of 3D reconstructed geophysical properties as probabilistic sampling constrain or guideline to speed up the sampling scheme.\par


The BO method as discussed here is not limited to geophysical inversion and drill-core sampling optimization alone, but can be applied to a wide range of sensor fusion problems and allocation problems with expensive cost functions for any new measurement (e.g., costly drill cores). This offers solutions for possible questions such as: Which additional sensor type would provide the maximum gain? What is the optimal mixture of sensors? What is the minimum amount of sensors to be activated given the total energy constraints and how should they be distributed? How to position sensor grids over time if the model state is dynamic or has a moving target? Future research may address these problems in more detail, ideally with a network of probabilistic modeling components where each node is capable of complete propagation of uncertainties from input node to output node.

\section{Conclusions}

One of the most important aspects when dealing with incomplete information is where to acquire new observations, in particular if measurements are very costly or resources are limited. The probabilistic framework presented here combines inverse modeling and multi-objective Bayesian optimisation to find the global optimal solution for allocating new measurement. The joint inversion model uses sparse Gaussian Process kernels and multi-linear forward models to construct 3D-geophysical properties from 2D-sensor data. One key advantage of this method is to solve simultaneously multiple sensor measurements by taking into account the correlations between distinct geological aspect, e.g., density and magnetic susceptibility. Furthermore, the solution of the inverse problem provides a complete posterior distribution for all geophysical properties under consideration at any location or grid resolution.\par

Multiple optimisation strategies have been evaluated on a set of synthetic and real geophysical data by addressing a specific example of a joint inverse problem, i.e., where to place new drill-core measurements given 2D gravity and magnetic sensor data, while taking into account potential drill-core costs and any pre-existing drill-core data. We conclude that the proposed method can accurately reconstruct 3D geophysical properties and their associated uncertainties, and significantly outperform traditional non-bayesian optimisation sampling schemes, which ultimately leads to more efficient sensor selection, higher information gain, and reduced cumulative costs. Moreover, we find that the model is flexible enough to accommodate a variety of prior knowledge, such as the characteristic length-scales of geological structures and the correlation coefficients between different geophysical properties. We propose that the same model can be applied to a wide range of sensor fusion problems and allocation problems - ranging from constraints limiting surface access for data collection to adaptive multi-sensor positioning.

\section{ACKNOWLEDGMENTS}

The authors would like to thank the anonymous reviewers for their valuable suggestions and comments to improve and clarify this manuscript. This research was supported by the Sydney Informatics Hub, a Core Research Facility of the University of Sydney.

\clearpage

\append{Gravity Forward Model}

\label{section_gravmodel}
The values of the matrix $G_{const}$ can be determined analytically for a prism by its origin and end coordinates $z^0$ and $z^e$, and the corresponding sensitivity at location x is given by \citep{nagy2000gravitational,reid2013bayesian}

\begin{equation}
T _ { \mathbf { x } , \mathbf { z } } = G_{const} g ( \mathbf { z } ) \mid^{  \overline { z } _ { 1 } ^ { e }}_{\overline { z } _ { 1 } ^ { o}} \mid^{\overline { z } _ { 2 } ^ { e }}_{\overline { z } _ { 2 } ^ { o}} \mid^{\overline { z } _ { 3 } ^ { e }}_{\overline { z } _ { 3 } ^ { o }},
\end{equation}
where $G_{const}$ is the gravitational constant, and 
\begin{eqnarray}
{ g ( \mathbf { z } ) = z _ { 1 } \log \left( z _ { 2 } + r \right) + z _ { 2 } \log \left( z _ { 1 } + r \right) - z _ { 3 } \arctan \frac { z _ { 1 } z _ { 2 } } { z _ { 3 } r } }; \\ { \overline { \mathbf { z } } ^ { o } = \mathbf { z } ^ { o } - \mathbf { x } ; \overline { \mathbf { z } } ^ { e } = \mathbf { z } ^ { e } - \mathbf { x } ; \text { and } r = \sqrt { z _ { 1 } ^ { 2 } + z _ { 2 } ^ { 2 } + z _ { 3 } ^ { 2 } } }. 
\end{eqnarray}

\bibliographystyle{seg}
\bibliography{refs}

\clearpage


\end{document}